\documentclass[prd,twocolumn,superscriptaddress,nofootinbib,10.5pt]{revtex4-1}  
\usepackage{epsfig}  
\usepackage{amsmath} 
\usepackage{amsfonts}   
\usepackage{float}    
\usepackage{amssymb}    
\usepackage{slashed} 
\usepackage{color}       
\usepackage{pbox}
\usepackage{subfigure}
\usepackage{array,multirow,makecell}
\usepackage[colorlinks,citecolor=blue, linkcolor = blue, urlcolor = blue]{hyperref}
\usepackage{tabularx}
\usepackage{titlesec} 
\usepackage{scrextend}
\usepackage{ulem}
\usepackage{natbib} 
\usepackage{array,multirow,makecell}
\usepackage{lipsum}
\usepackage{tikz,xcolor,hyperref}

\def\lsim{\raise0.3ex\hbox{$\;<$\kern-0.75em\raise-1.1ex\hbox{$\sim\;$}}}
\def\gsim{\raise0.3ex\hbox{$\;>$\kern-0.75em\raise-1.1ex\hbox{$\sim\;$}}}

\newcommand{\be}{\begin{equation}}
\newcommand{\ee}{\end{equation}}
\newcommand{\bea}{\begin{eqnarray}}
\newcommand{\eea}{\end{eqnarray}}

\definecolor{lime}{HTML}{A6CE39}

\newcommand{\AddrIOP}{
	Institute of Physics, Bhubaneswar, Sachivalaya Marg, Sainik School, Bhubaneswar 751005, India}

\graphicspath{{Figures/}}
\begin{document}

\hfill \preprint~KIAS-P22070

\title{CMB imprints of high scale non-thermal leptogenesis}

\author{Anish Ghoshal}
\email{anish.ghoshal@fuw.edu.pl}
\affiliation{Institute of Theoretical Physics, Faculty of Physics, University of Warsaw,ul.  Pasteura 5, 02-093 Warsaw, Poland}
\author{Dibyendu Nanda}
\email{dnanda@kias.re.kr}
\affiliation{School of Physics, Korea Institute for Advanced Study, Seoul 02455, South Korea}
\author{Abhijit Kumar Saha}
\email{psaks2484@iacs.res.in}
\email{abhijit.saha@iopb.res.in}
\affiliation{School of Physical Sciences, Indian Association for the Cultivation of Science,\\ 2A $\&$ 2B Raja S.C. Mullick Road, 
Kolkata 700032, India}
\affiliation{\AddrIOP}

\begin{abstract}
We study the imprints of high scale non-thermal leptogenesis on cosmic microwave background (CMB) from the measurements of inflationary spectral index ($n_s$) and tensor-to-scalar ratio ($r$), which otherwise is inaccessible to the conventional laboratory experiments. We argue that non-thermal production of baryon (lepton) asymmetry from subsequent decays of inflaton to heavy right-handed neutrinos (RHN) and RHN to SM leptons is sensitive to the reheating dynamics in the early Universe after the end of inflation. Such dependence provides detectable imprints on the $n_s-r$ plane which is well constrained by the Planck experiment. We investigate two separate cases, (I) inflaton decays to radiation dominantly and (II) inflaton decays to RHN dominantly which further decays to the SM particles to reheat the Universe adequately. Considering a class of $\alpha-$ attractor inflation models, we obtain the allowed mass ranges for RHN for both cases and thereafter furnish the estimates for $n_s$ and $r$. The prescription proposed here is general and can be implemented in various kinds of single-field inflationary models given the conditions for non-thermal leptogenesis are satisfied. 

\end{abstract}

\maketitle

\section{Introduction} 
\label{sec:intro}
Observation of neutrino oscillations at various neutrino reactor experiments manifests that neutrinos are massive and have non-zero mixings \cite{Super-Kamiokande:2001bfk,Super-Kamiokande:2002ujc,Super-Kamiokande:2005mbp,SNO:2002tuh,KamLAND:2008dgz,T2K:2011ypd,DoubleChooz:2011ymz,T2K:2013ppw}. On the cosmological front, the BBN (Big bang nucleosynthesis), CMBR (Cosmic microwave background radiation) and LSS (Large scale structure) measurements favor neutrino mass to remain in the sub-eV range. The simplest mechanism to fit the neutrino oscillation data and explain the origin of the Standard Model (SM) neutrino masses is the type-I seesaw mechanism \cite{Minkowski:1977sc,Mohapatra:1979ia,Yanagida:1979as,Gell-Mann:1979vob} where the SM is extended with three right handed neutrinos (RHN), singlet under SM gauge symmetry. Remarkably, such minimal extension can also explain the cosmic matter-antimatter asymmetry which is dubbed as the baryon asymmetry of the Universe (BAU). The RH neutrinos present in the type-I seesaw model \cite{Mohapatra:1979ia,Minkowski:1977sc} with lepton number violating (LNV) Majorona mass are inherently unstable and decay to SM Higgs plus leptons. This out-of-equilibrium decay process at 1-loop picks up CP violation (CPV) via the complex Yukawa couplings, leading to asymmetric decay into the leptons than in the anti-lepton counterpart \cite{Fukugita:1986hr,Luty:1992un,Plumacher:1996kc,Covi:1996wh}. Thus the three Sakharov conditions being fulfilled, we can have successful baryogenesis in the early universe when afterwards the lepton asymmetry partially gets converted to the positive baryon asymmetry that we observe today\,\cite{Fukugita:1986hr,Luty:1992un,Plumacher:1996kc,Covi:1996wh}.

The production of lepton asymmetry at the early stages of the Universe can be thermal \cite{Fukugita:1986hr,Giudice:2003jh} or non-thermal \cite{Lazarides:1990huy,Murayama:1992ua,Kolb:1996jt,Giudice:1999fb,Asaka:1999yd,Asaka:1999jb,Jeannerot:2001qu,Hamaguchi:2001gw,Asaka:2002zu,Senoguz:2007hu,Fukuyama:2005us,Endo:2006nj,Hahn-Woernle:2008tsk,Co:2022bgh,Barman:2022qgt} in nature. In the case of thermal leptogenesis, the reheating temperature has to be larger than the RH neutrino mass scale ($M_N<T_R$) such that a non-zero initial abundance of the RH neutrino can be created efficiently from the thermal bath. In the non-thermal case, the condition $M_N<T_R$ is not a necessity. The required initial abundance of RHN can be alternatively created non-thermally from a heavy scalar decay present in the early Universe. The scalar field can be identified with the inflaton which leads to an accelerated expansion at the beginning of the universe, in order to solve the
horizon and the flatness problems. The same field could also be responsible for the quantum generation of the primordial fluctuations seeding the large scale structure (LSS) of the Universe (see \cite{Martin:2013tda} for a review).

Despite the elegant explanation for the tiny SM neutrino masses and the generation of matter-antimatter asymmetry, the seesaw mechanism is excruciatingly difficult to test in laboratories, since in order to successfully drive leptogenesis, the right-handed neutrino mass scale has to be above $ \gtrsim  10^{9}$~GeV (see, {\it e.g.}\/, \cite{Buchmuller:2004nz})\footnote{This bound can be evaded in case of resonant production of lepton asymmetry in presence of nearly degenerate RH neutrino species \cite{Pilaftsis:2003gt}} \footnote{With some fine tuning, it is also possible to lower the scale of the non-resonant thermal leptogenesis to as low as $10^6$ GeV \cite{Moffat:2018wke}.}. The indirect tests for high scale leptogenesis, of course, exist based on neutrino-less double beta decay and lepton flavor and CP violating decays of mesons \cite{DellOro:2016tmg}, via CP violation in neutrino oscillation \cite{Endoh:2002wm,Esteban:2016qun}, by the structure of the mixing matrix \cite{Bertuzzo:2010et}, or from theoretical constraints stemming from the demand of Higgs vacuum remaining meta-stable in the early universe \cite{Ipek:2018sai,Croon:2019dfw}.
And more recently Gravitational Waves (GW) of primordial origins like that from cosmic strings \cite{Dror:2019syi}, domain walls \cite{Barman:2022yos}, nucleating and colliding vacuum bubbles \cite{Dasgupta:2022isg,Borah:2022cdx} or other topological defects \cite{Dunsky:2021tih} and primordial blackholes \cite{Bhaumik:2022pil} have been proposed to constrain as well shed some light on high-scale leptogenesis scenarios. Under these circumstances, it is necessary, although highly challenging to find new and complementary tests of such heavy neutrino sectors and consequently the leptogenesis mechanism.

Motivated by this, in this paper, we envisage the scope of tracing the fingerprints of high scale non-thermal leptogenesis at CMB experiments.
If the lepton asymmetry is produced via the transfer of energy density from inflaton sector to the lepton sector, then the amount of final lepton asymmetry yield is dependent on the reheating history of the Universe. On the other hand, for a given a model of inflation the predictions of inflationary observables namely the spectral indices and tensor to scalar ratio also influenced by the post-inflationary \,\cite{Drewes:2015coa} physics {\it e.g.} number of e-folds during reheating era as speculated in ref.\,\cite{Drewes:2015coa} and subsequently studied in details in Refs. \cite{Drewes:2017fmn,Drewes:2019rxn,Drewes:2022nhu}.  The aforementioned two observations suggest that the non-thermal leptogenesis at early Universe is expected to leave non-negligible imprints in the CMB predictions for inflationary observables which we pursue in this paper\footnote{The impact of dark matter production at the very early Universe on the inflationary observables have been studied by the authors of \cite{Maity:2018dgy,Maity:2018exj,Haque:2020zco}.}.

We revisit the inflatonary reheating epoch \cite{Kofman:1994rk,Kofman:1997yn} in $\alpha$-attractor models of inflation \cite{Kallosh:2013lkr,Kallosh:2013hoa,Kallosh:2013yoa,Kallosh:2013pby,Kallosh:2013maa,Galante:2014ifa}. Interestingly, when we take into account the non-thermal production of RHN (that subsequently yields lepton asymmetry) from tree level inflaton decay, the reheating temperature of the Universe cannot be arbitrary, rather guided by the observed amount of baryon asymmetry of the Universe. This in turn provides a distinct prediction of the non-thermal leptogenesis on the $n_s-r$ plane, which turns out to be stronger than the one provided by PLANCK/BICEP experiment \cite{BICEP:2021xfz}. 

In our analysis, we have assumed that the inflaton decays perturbatively. We have separately discussed two possible sub-cases: (I) inflaton decays dominantly to radiation and (II) tree level interaction between inflaton and radiation is absent. In the latter case, the reheating of the Universe is realized solely from the decay of RHNs. We have followed a detailed numerical approach considering the finite epoch of the perturbative reheating era. We point out the allowed ranges of RHN mass in order to realize the non-thermal leptogenesis for both Case I and Case II.  {From these two case studies, we have found that the occurrence of a successful non-thermal leptogenesis scenario indeed imposes stringent and a generic restriction in the $n_s-r$ plane, irrespective of whether a radiation dominated Universe is obtained from inflaton or RHN decay.}

\section{$\alpha$-attractor inflation model}
A general form of the $\alpha$-attractor inflaton potential (known as E model) is read as \cite{Kallosh:2013maa},
\begin{align}
 V(\phi)=\Lambda^4 \left(1-e^{-\sqrt{\frac{2}{3\alpha}}\frac{\phi}{M_P}}\right)^{2n}\label{eq:InfPot},
\end{align}
where $M_P$ stands for the reduced Planck scale and $\Lambda$ represents a mass scale that determines the energy scale of the inflation. A special case of Eq.(\ref{eq:InfPot}) with $\alpha=1$ and $n=1$ mimics the standard Higgs-Starobinsky inflaton potential \cite{Bezrukov:2007ep}. We first calculate the spectral index ($n_s$) under the slow roll approximation to express it in terms of the model parameters  $\alpha$ and $n$.
\begin{align}
&n_s=1-\frac{8n\left(e^{\sqrt{\frac{2}{3\alpha}}\frac{\phi_k}{M_P}}+n\right)}{3\alpha\left(e^{\sqrt{\frac{2}{3\alpha}}\frac{\phi_k}{M_P}}-1\right)^2},
\end{align}
where $\phi_k$ is the inflaton field value at horizon exit. This can be simply translated to write $\phi_k$ as a function of $n_s$.
\begin{align}
~\phi_k=\sqrt{\frac{3\alpha}{2}}\,M_P\,{\rm ln}\left(1+\Delta(n_s)\right) \label{eq:phiK},
\end{align}
where $\Delta(n_s)=\frac{4 n+\sqrt{16 n^2+24\alpha n(1-n_s)(1+n)}}{3\alpha(1-n_s)}$. Next, we make an estimate for the inflaton field value at the end of inflation by equating one of the slow roll parameters (max[$\epsilon,\eta$]) to unity as given by,
\begin{align}
\phi_{\rm end }=\sqrt{\frac{3\alpha}{2}}\,M_P\,{\rm ln}\left(\frac{2n}{\sqrt{3\alpha}}+1\right).\label{eq:phiEND}    
\end{align}
Utilizing Eq.(\ref{eq:phiK}) and Eq.(\ref{eq:phiEND}), we compute quantities namely tensor to scalar ratio and the number of e-fold analytically which are given by,
\begin{align}
  r&=\frac{64 n^2}{3\alpha\left(e^{\sqrt{\frac{2}{3\alpha}}\frac{\phi_k}{M_P}}-1\right)^2}\nonumber\\
  &=\frac{192\alpha n^2(1-n_s)^2}{\left[4 n+\sqrt{16 n^2+24\alpha n(1-n_s)(1+n)}\right]^2},\label{eq:r}\\
  N_k&=\frac{3\alpha}{4 n}\left[e^{\sqrt{\frac{2}{3\alpha}}\frac{\phi_k}{M_P}}-e^{\sqrt{\frac{2}{3\alpha}}\frac{\phi_{\rm end}}{M_P}}-\sqrt{\frac{2}{3\alpha}}\frac{(\phi_k-\phi_{\rm end})}{M_P}.\right]\label{eq:Nk}
\end{align}
The scalar potential at the end of inflation and the scalar perturbation spectrum are obtained as,
\begin{align}
& V_{\rm end}= \Lambda^4\left(\frac{2n}{2n+\sqrt{3\alpha}}\right)^{2n},\label{eq:Vend}\\
 & A_s=\frac{3 V(\phi_k)}{4\pi^2r}.
\end{align}
The observed value of $A_s^{\rm obs}=2.2\times 10^{-9}$ \cite{Planck:2018vyg} precisely fixes one of the model parameters $\Lambda$
as,
\begin{align}
    \Lambda=& M_P\left(\frac{3\pi^2 r A_s^{\rm obs}}{2}\right)^{1/4}\nonumber\\
    & \times\left[\frac{2n(1+2n)+\sqrt{4 n^2+6\alpha(1+n)(1-n_s)}}{4n(1+n)}\right]^{n/2}.
\end{align}
It is worth mentioning that each of the quantities ($r,N_k,V_{\rm end} {~\rm and~} \Lambda$) have been expressed exclusively in terms of three model parameters $(n,\alpha$ and $n_s)$. This will be immensely useful in connecting the lepton asymmetry from the RHN decay to the inflationary observables as we will describe in a while.

\section{Non-thermal Leptogenesis from inflaton decay}

To embed the non-thermal leptogenesis in the inflationary framework we propose the following Lagrangian in a model independent manner:
\begin{align}
-\mathcal{L}\supset & y_N\,\phi\,\overline{N^C}\,N+ y_R\,\phi\, \overline{X}\,X+Y_\nu \overline{l_L}\widetilde{H}N \\ &~~~~~~~~~~~~~~~~~~~~~+ \frac{M_N}{2} \overline{N^C} N + h.c.,\label{eq:newL}
\end{align}
 where $X$ represents a Dirac fermion which is part of the radiation bath after the completion of reheating era \footnote{\color{black}Here we remain agnostic about an UV complete framework and only consider $X$ to be a sample particle which is part of radiation bath. As an instance one can consider an $U(1)_X$ gauged extension of the SM where $X$, being a gauge singlet vector fermion is coupled to SM sector via $U(1)_X$ gauge boson and having same temperature as of SM radiation bath.}. The $N$ (assumed to be of Majorona nature and having a bare mass $M_N$) is the RH neutrino which decays to SM leptons ($l_{L}$) and Higgs ($H$) and subsequently contribute to the yield of lepton asymmetry. We assume all the coupling coefficients $y_N$,\,$y_R$ real and positive. The neutrino Yukawa coupling $Y_\nu$ could be complex and source the CP violation in the SM lepton sector. In the present analysis we deal with sufficiently small Yukawa couplings $y_N$ and $y_R$ such that they do not disturb the shape of the inflationary potential through radiative corrections.

The first term in Eq.(\ref{eq:newL}) leads to the non-thermal production of $N$ from tree level inflaton decay. Whereas, the second term is responsible for inflaton decay to radiation at tree level. The third term triggers the out-of-equilibrium decay of RHN to SM leptons. In principle it is essential to add at least two SM gauge singlet RH Majorona neutrinos in order to satisfy the neutrino oscillation data \cite{ParticleDataGroup:2020ssz}) in the usual type-I seesaw framework. We consider here $N_{2}$ being larger than the inflaton mass and hence their production from inflaton decay is expected to be suppressed. On the other hand, $N_1$ is considered to be lighter than the inflaton and can be produced efficiently.
In view of this, we ignore the dynamics of $N_2$
and track the evolution of $N_1$ only which is same as $N$ in Eq.(\ref{eq:newL}).

As mentioned in the introduction section, we will examine two distinct cases: {\bf (I)} inflaton dominantly decays to radiation and reheats the Universe {\it i.e.} ${\rm Br_{\phi\to X X}}> {\rm Br}_{\phi\to N N}$ and {\bf (II)} inflaton decays to RHN first and its further decay to $SU(2)_L$ doublets reheats the Universe which implies $y_R\ll y_N$. 
{\color{black}In the first case, we also need to ensure that the produced RH neutrino is not long-lived and radiation energy density ($\rho_R$) always remains much larger than the energy density of RH neutrino ($\rho_N$) such that the RHN never dominates the energy budget of the Universe. In that case, the decay of RHN contributes negligibly to the $\rho_R$ and the comoving entropy density $s a^3$ remains almost constant before and after the decay of $N$. Thus the reheating temperature of the Universe is controlled by the inflaton decay to radiation only.} For simplicity, in the present work we have assumed that both the inflaton and RHN decay perturbatively. 

We define the decay widths of $\phi$ to both radiation and $N$ as given by \cite{Hahn-Woernle:2008tsk},
\begin{align}
    \Gamma_{\phi}^N=\frac{|y_N|^2m_\phi}{16\pi}\left(1-\frac{4 M_N^2}{m_\phi^2}\right)^{3/2},~~~\Gamma_{\phi}^R\simeq\frac{|y_R|^2m_\phi}{8\pi},
\end{align}
while the decay width for the RHN is $\Gamma_N\simeq \frac{|Y_\nu|^2 M_N}{16\pi}$. In the above, we have considered the SM particles to be much lighter than the inflaton. The set of Boltzmann equations that govern the evolution of energy densities of various species, number densities for $N$ and the yield of lepton asymmetry is given by \cite{Hahn-Woernle:2008tsk}:
\begin{align}
&\frac{d\rho_\phi}{dt}+3H(p_\phi+\rho_\phi)=-\Gamma_\phi^N\rho_\phi-\Gamma_\phi^R\rho_\phi,\label{eq:leprhoPh}\\
&\frac{d\rho_R}{dt}+3H(p_R+\rho_R)=\Gamma_\phi^R\rho_\phi+\Gamma_{N}\rho_N,\\
&\frac{d\rho_N}{dt}+3H(p_N+\rho_N)=\Gamma_\phi^N\rho_\phi-\Gamma_N\rho_N,\\
&\frac{dn_{B-L}}{dt}+3 Hn_{B-L}=-\frac{\varepsilon \rho_N\Gamma_N}{M_N}\label{eq:lepAs},
\end{align}
where $\varepsilon$ is the CP asymmetry parameter {\color{black}which is a function of the neutrino Yukawa couplings and RHN mass scales (see
 Eq.(\ref{eq:assyP})). We have evaluated the neutrino Yukawa couplings using the Casas Ibara (CI) parametrisation \cite{Casas:2001sr} using the best-fit values of the neutrino oscillation parameters. This helps us further to write the lepton asymmetry parameter $\varepsilon$ as a function of $M_N$ and the complex rotational angle, $\theta$ appearing in the CI parametrisation.} In the Boltzmann equation for $n_{B-L}$, we have ignored the washout effects since it is suppressed in the case of non-thermal production of lepton asymmetry which we show later on. The lepton asymmetry as obtained from Eq.(\ref{eq:lepAs}) can be converted to the baryon  asymmetry induced by the \textit{sphaleron} process before electroweak Phase transition (EWPT) as given by \cite{Buchmuller:2004nz},
\begin{align}
 \frac{n_B}{s}\sim  \left(\frac{8}{23}\right)\times\frac{n_{B-L}}{s}~. \label{eq:nbs}
\end{align}
The experimentally observed value of $\frac{n_B}{s}$ is $8.7\times 10^{-11}$ \cite{Planck:2018vyg}.

\section{CMB imprints of non-thermal leptogenesis}
{\color{black}Some of the earlier studies on non-thermal leptogenesis rely on the approximate analytical formulation of final baryon asymmetry abundance. It is found that the amount of lepton asymmetry yield from inflaton decay is proportional to the reheating temperature of the Universe and the branching ratio of the inflaton decay to RH neutrinos as depicted by the relation, \cite{Giudice:1999fb,Asaka:1999yd,Antusch:2010mv,Antusch:2018zvu}: 
\begin{align}
    \frac{n_{B-L}}{s} \approx \frac{3}{2}\times{\rm Br_{\phi\to N N}}\times \left(\frac{\varepsilon T_R}{m_\phi}\right)\label{eq:anaNL}
\end{align}
Note that, Eq.(\ref{eq:anaNL}) holds under the approximation of instantaneous inflaton decay. In an attempt to deliver an accurate description of particle production during the reheating era for both Case I and Case II, we follow the numerical approach by properly solving the relevant set of Boltzmann equations in this work considering a finite epoch of the reheating period.} Additionally, it is appraised earlier \cite{Garcia:2020eof,Giudice:2000ex} that the maximum temperature of the Universe ($T_{\rm max}$) turns out to be always a few order larger than the original reheating temperature ($T_{\rm RH}$) of the Universe under the non-instantaneous perturbative reheating consideration. Therefore, a more relevant condition for the non-thermalization of the RHN in the early Universe would be $M_N>T_{\rm max}$ \cite{Datta:2022jic} instead of $M_N>T_{\rm RH}$. This fact further enlightens the importance of solving the Boltzmann equations Eqs.\,(\ref{eq:leprhoPh}-\ref{eq:lepAs}) for particle production considering a finite reheating epoch \footnote{For an analytical estimate of $T_{\rm max}$ see \cite{Kolb:2003ke}.} \footnote{Once the RH neutrinos abundance is formed the inflaton later on can also mediate leptogenesis via higher-dimensional operators involved  \cite{Hamada:2015xva,Hamada:2016npz}}.

Considering FRW ansatz, the e-folding number from the end of inflation to the end of the reheating epoch is written as
\begin{align}
 N_{\rm re}={\rm ln}\left(\frac{a_{\rm re}}{a_{\rm end}}\right)=-\frac{1}{3(1+\overline{\omega}_{\rm re})}{\rm ln}\left(\frac{\rho_{\rm re}}{\rho_{\rm end}}\right),\label{eq:Nre}
\end{align}
where $a_{\rm end}$ and $a_{\rm re}$ are the corresponding scale factors at the end of inflation and end of reheating respectively having respective energy densities, $\rho_{\rm re}$ and $\rho_{\rm end}$. The quantity $\rho_{\rm re}$ can be computed numerically by solving the set of Boltzmann equations (see Eqs.\,(\ref{eq:leprhoPh}-\ref{eq:lepAs})). In Eq.(\ref{eq:Nre}), the averaged equation state from the end of inflation to the end of reheating is indicated by $\overline{\omega}_{\rm re}$ which is defined as,
\begin{align}
 \overline{\omega}_{\rm re}=\frac{1}{N_{\rm re}}\int_0^{N_{\rm re}}\omega(N_e)\,dN_e.
\end{align}
It is shown in \cite{Lozanov:2016hid}, that the equation of state parameter during oscillation era is approximately $\overline{\omega}_{\rm re} \approx \frac{n-1}{n+1}$ for a generic inflationary potential $V(\phi)\propto \phi^n$ \footnote{In $\alpha$-attractor models of inflation, the inflation potential mimics the form $V(\phi)\propto \phi^n$ during reheating epoch.}.
Now at horizon exit $k=a_k\,H_k$, and one can write
\begin{align}
 {\rm ln}\left(\frac{k}{a_kH_k}\right)={\rm ln}\left(\frac{a_{\rm end}}{a_k}\,\frac{a_{\rm re}}{a_{\rm end}}\,\frac{a_0}{a_{\rm re}}\,\frac{k}{a_0 H_k}\right)=0,
\end{align}
where $a_0$ stands for the scale factor at the present time. This further implies,
\begin{align}
 N_k+N_{\rm re}+{\rm ln}\left(\frac{a_0}{a_{\rm re}}\right)+{\rm ln}\left(\frac{k}{a_0 H_k}\right)=0.\label{eq:totEV}
\end{align}
Next task is to express the ratio $\frac{a_{\rm re}}{a_0}$ as function of $N_{\rm re}$. Using the entropy conservation principle between the end of reheating and the present epoch, we find,
\begin{align}
 \frac{a_{\rm re}}{a_0}=\left(\frac{43}{11 g_{*s}}\right)^{1/3}\left(\frac{\pi^2g_*T_0^4}{30\rho_{\rm re}}\right)^{1/4},
 \label{eq:scalefac1}
\end{align}
 Further simplification of Eq.(\ref{eq:scalefac1}) is possible by connecting $\rho_{\rm re}$ to $\rho_{\rm end}$ (or $V_{\rm end}$) by Eq.(\ref{eq:Nre}). Now during inflation,
 \begin{align}
  \rho=\frac{1}{2}\dot{\phi}^2+V(\phi). \label{eq:energyD}
 \end{align}
According to Klein Gordan equation (during inflation),
\begin{align}
 \ddot{\phi}+3 H\dot{\phi}+V^\prime(\phi)=0.
\end{align}
Assuming $\ddot{\phi}\ll 3 H\dot{\phi},V^\prime(\phi)$
and $H^2\simeq\frac{V(\phi)}{3 M_P^2}$, we find from Eq.(\ref{eq:energyD}),
\begin{align}
 \rho\simeq V(\phi)\left(1+\frac{\epsilon}{3}\right).
\end{align}
Hence 
\begin{align}
 \rho_{\rm end}\simeq \frac{4}{3}V_{\rm end}.
\end{align}
with $\epsilon\sim 1$ near the end of inflation.  We find the simplified relation between $\rho_{\rm re}$ and $\rho_{\rm end}$ as can be written as (following Eq.(\ref{eq:Nre}))
\begin{align}
 \rho_{\rm re}=\frac{4}{3} V_{\rm end}e^{-3N_{\rm re}(1+\overline{\omega}_{\rm re})}
\end{align}
With this Eq.(\ref{eq:scalefac1}) can be translated into,
\begin{align}
 {\rm ln}\left(\frac{a_{\rm re}}{a_0}\right)=&\frac{1}{3}{\rm ln}\left(\frac{43}{11 g_{s*}}\right)+\frac{1}{4}\left(\frac{\pi^2g_*}{30}\right)+\frac{1}{4}{\rm ln}\left(\frac{3 T_0^4}{4 V_{\rm end}}\right)\nonumber\\
 & \hspace{3cm} +\frac{3 N_{\rm re}(1+\overline{\omega}_{\rm re})}{4}.\label{eq:area0}
\end{align}

We replace Eq.(\ref{eq:area0}) in Eq.(\ref{eq:totEV}) and use $H_k=\frac{\pi M_P\sqrt{r A_s}}{\sqrt{2}}$ to obtain,
\begin{align}
 N_{\rm re}=&\frac{4 }{3 \omega_{\rm re}-1}\Bigg[N_k+{\rm ln}\left(\frac{k}{a_0T_0}\right)+\frac{1}{4}{\rm ln}\left(\frac{40}{\pi^2g_*}\right)\nonumber\\
 &\hspace{2cm}+\frac{1}{3}{\rm ln}\left(\frac{11 g_{s*}}{43}\right)-\frac{1}{2}{\rm ln}\left(\frac{\pi^2 M_P^2\,r\,A_s}{2 V_{\rm end}^{1/2}}\right)\Bigg].\label{eq:Nre-inf}
\end{align}
\begin{figure}[h]
    \centering
    \includegraphics[height=6.7cm,width=8.3cm]{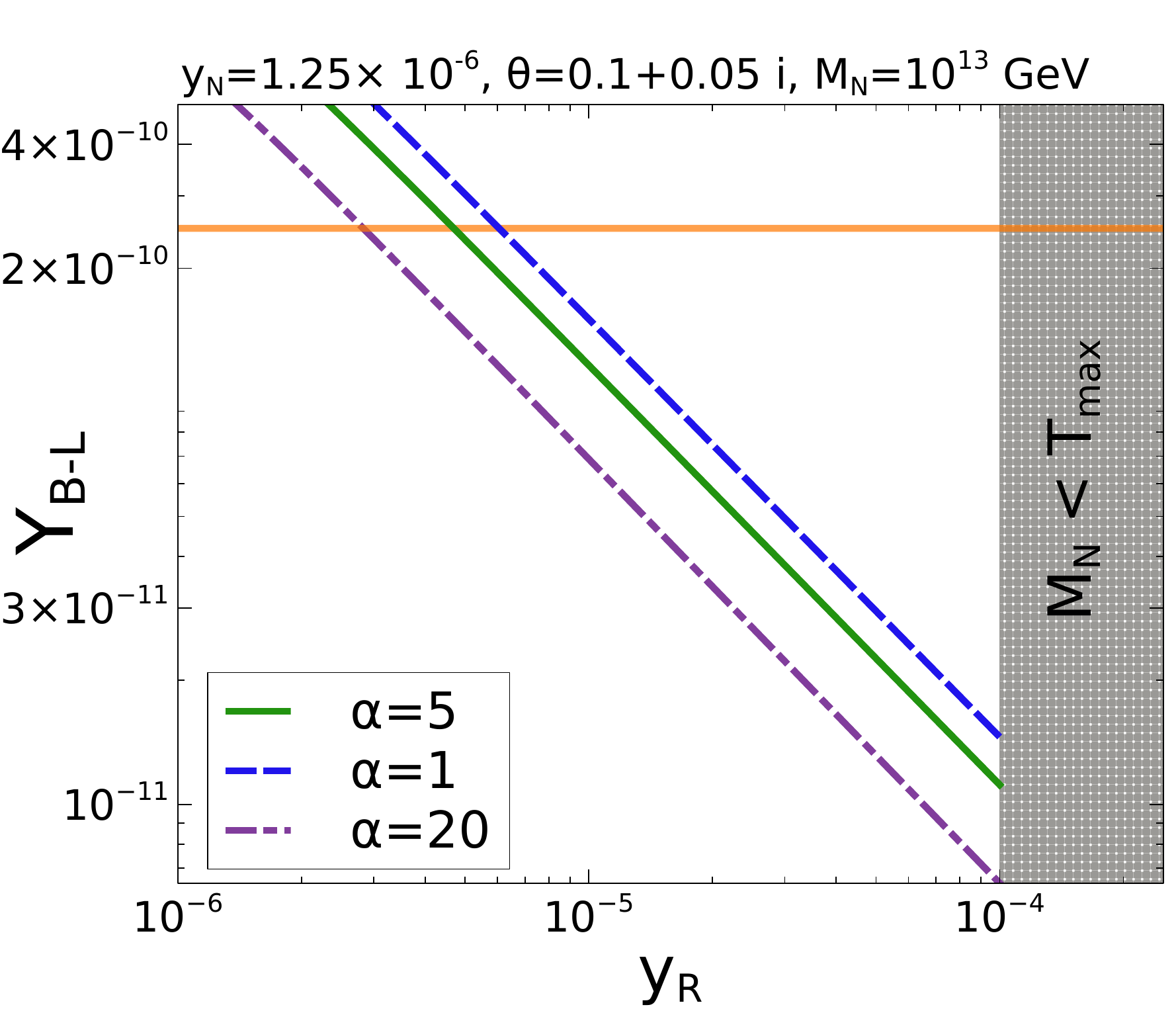}\\
    \includegraphics[height=6.7cm,width=8.3cm]{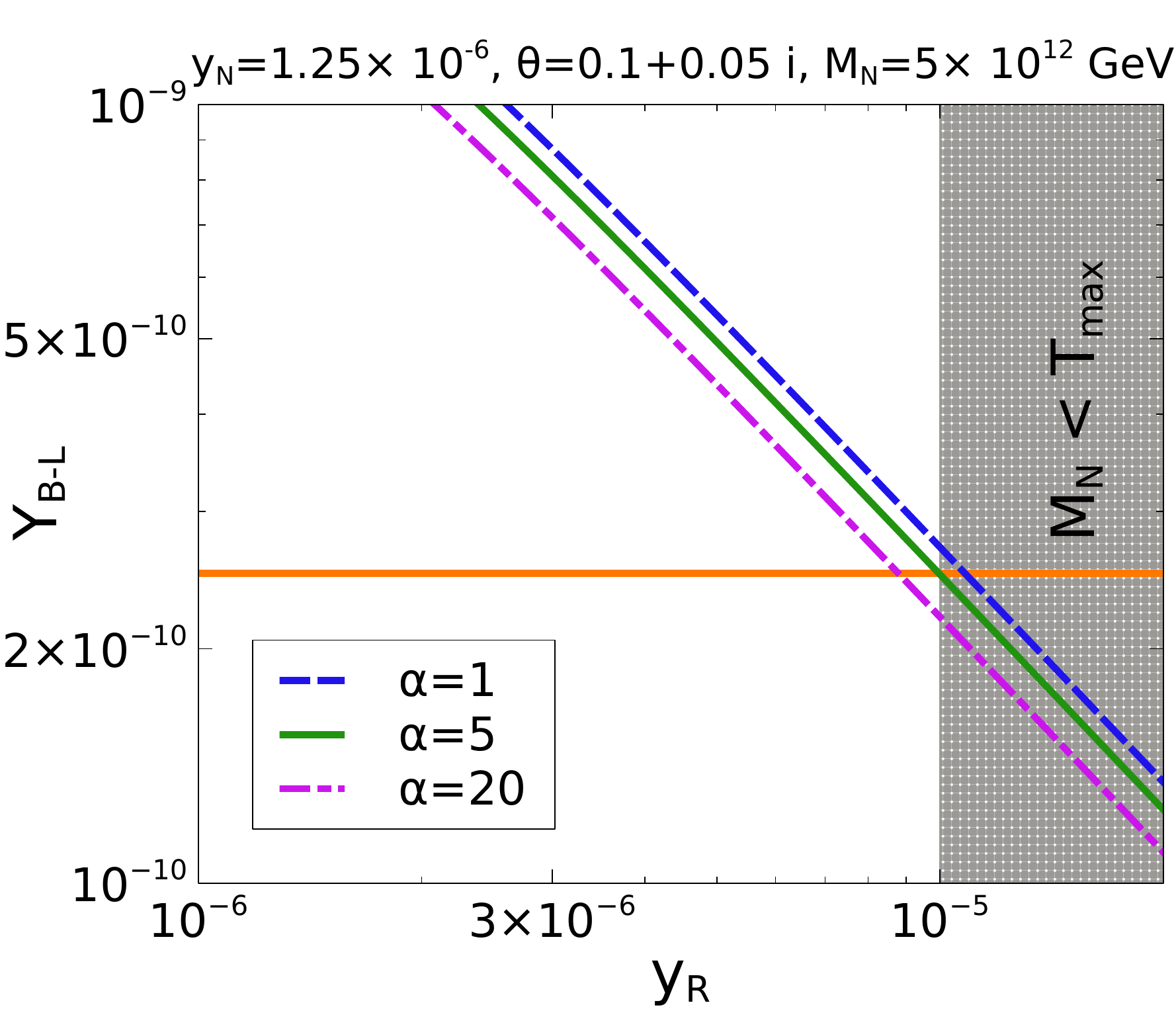}
    \caption{\color{black}\it Case I: The $B-L$ yield is plotted as a function of $y_R$ for different values of $\alpha$ and fixed $y_N\,,\theta$. The orange band represents the observed value of $Y_{B-L}=\frac{n_{B-L}}{s}$. We consider $M_N=10^{13}$ GeV and $5\times 10^{12}$ GeV in the top and bottom panels respectively.}
    \label{fig:y2ybl}
\end{figure}
\noindent The RHS of Eq.(\ref{eq:Nre-inf}) is purely a function of $n_s$ once we fix $\alpha,n$ and $A_s=A_s^{\rm obs}$ and replace $r,N_k,V_{\rm end}$ accordingly following Eqs.(\ref{eq:r}),\,(\ref{eq:Nk}) and (\ref{eq:Vend}) respectively. On the other hand, the LHS of Eq.(\ref{eq:Nre-inf}), carries the information of the duration of the reheating epoch, is a function of the radiation energy density after the completion of the reheating and can be obtained by solving Eqs.\,(\ref{eq:leprhoPh}-\ref{eq:lepAs}). Thus, Eq.(\ref{eq:Nre-inf}) manifests the clear dependence of $n_s$ (and $r$) on the $N_{\rm re}$ which is solely determined by the time required for the complete transfer of inflaton energy density to the radiation bath during reheating.

Interestingly, in the case of non-thermal leptogenesis, the RHN neutrino production and subsequently the baryon asymmetry of the Universe also rely on the reheating dynamics as evident from the coupled Boltzmann equations in Eqs.\,(\ref{eq:leprhoPh}-\ref{eq:lepAs})). These two observations open up the scope of understanding the dependence of lepton (baryon) asymmetry of the Universe on the inflationary observables $n_s$ (and $r$) on one hand and to probe the scale of leptogenesis on the other. The present study is aimed to examine such an intriguing connection quantitatively. To be specific, we attempt to find out whether the requirement of producing the observed amount of baryon asymmetry can put further constraints in the $(n_s,r)$ plane as already restricted by the Planck/BICEP data \cite{BICEP:2021xfz}.

The lepton asymmetry parameter $\varepsilon$ and decay width of RHN, $\Gamma_N$ is a function of neutrino sector parameters namely $Y_\nu$,\,$M_N$ {\it etc.}. The estimate of $Y_\nu$ as a function of $M_N$ is obtained by Casas Ibara (CI) parametrisation \cite{Casas:2001sr}, obeying the neutrino oscillation data. Note that in the conventional CI parametrisation in the presence of a $2\times 2$ RHN mass matrix one introduces a complex orthogonal matrix with complex rotational angle $\theta$ to achieve the desired structure of $Y_\nu$ (see appendix\,\ref{sec:CI} for details). {The complex rotational angle $\theta$ acts a source of the CP violation in the theory \footnote{The Dirac CP violation ($\delta_{\rm CP}$) plays no role in unflavored leptogenesis.}. When, $\theta=0$, indicating no CP violation in the theory, the lepton asymmetry yield will be zero. In principle, the complex rotational angle $\theta$ can take any value. In this work, since our primary objective is to provide a concrete prediction of $n_s$ and $r$ in a non-thermal leptogenesis framework while performing a broad scan we will vary the CP violating parameter $\theta$ randomly. However a large value of $\theta$ can cause a violation of the perturbativity limit ($Y_\nu<4\pi$) and at the same time induce electroweak vacuum instability which imposes the most stringent bound $Y_\nu^2\lesssim 0.5$\,\cite{Bambhaniya:2016rbb}. Therefore in the present analysis, we will restrict ourselves below a threshold value of $\theta$ where the electroweak vacuum stability bound remains unviolated.}
We have also considered $n=1$ in the inflationary potential as it gives rise to a bare mass term for the inflaton field (during reheating) which is $m_\phi^2=\frac{4\Lambda^4}{3\alpha M_P^2}$.  
\begin{figure*}[t]
    \centering
    \includegraphics[height=6.8cm,width=7.8cm]{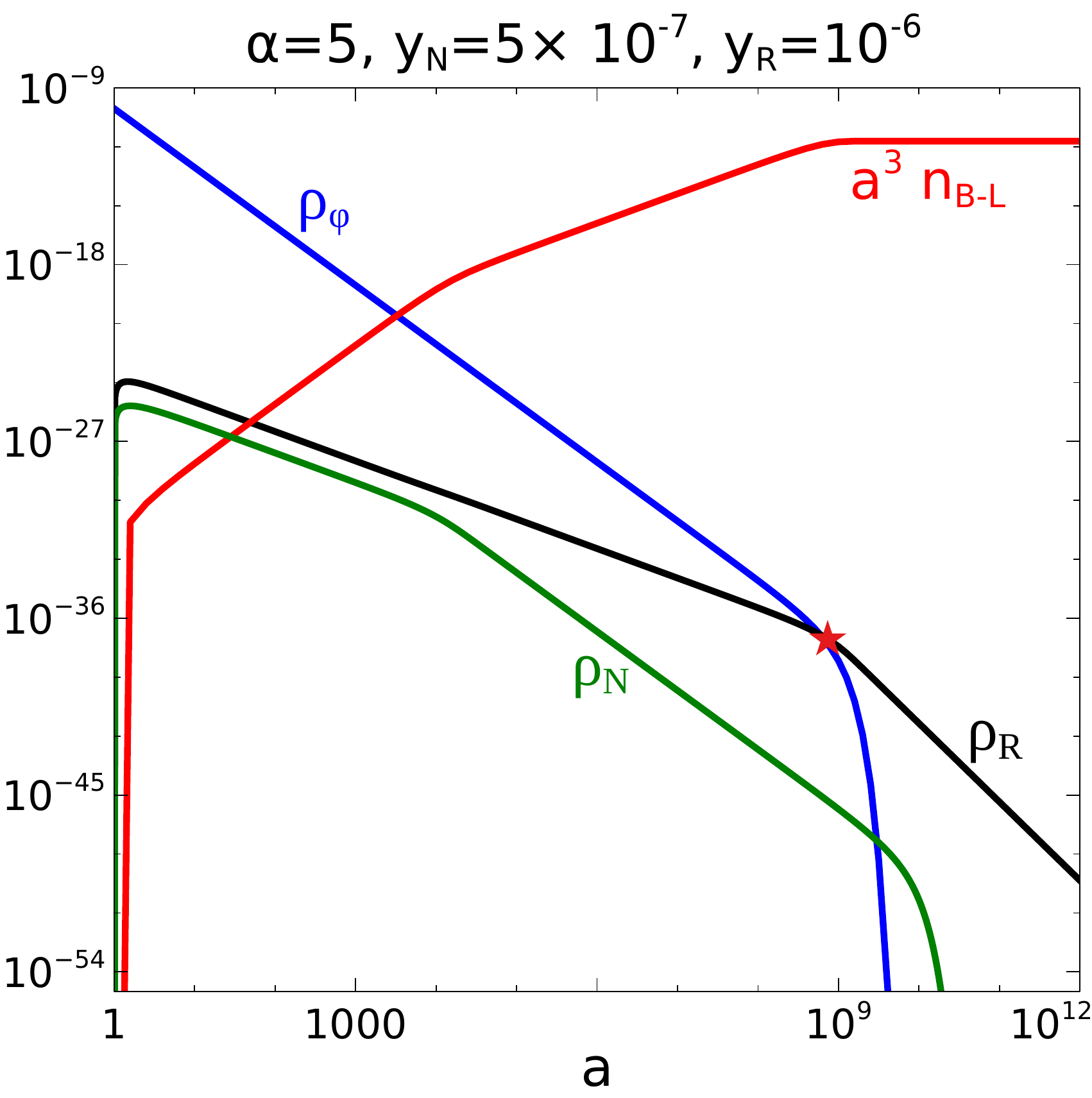}~~~
    \includegraphics[height=6.8cm,width=8.0cm]{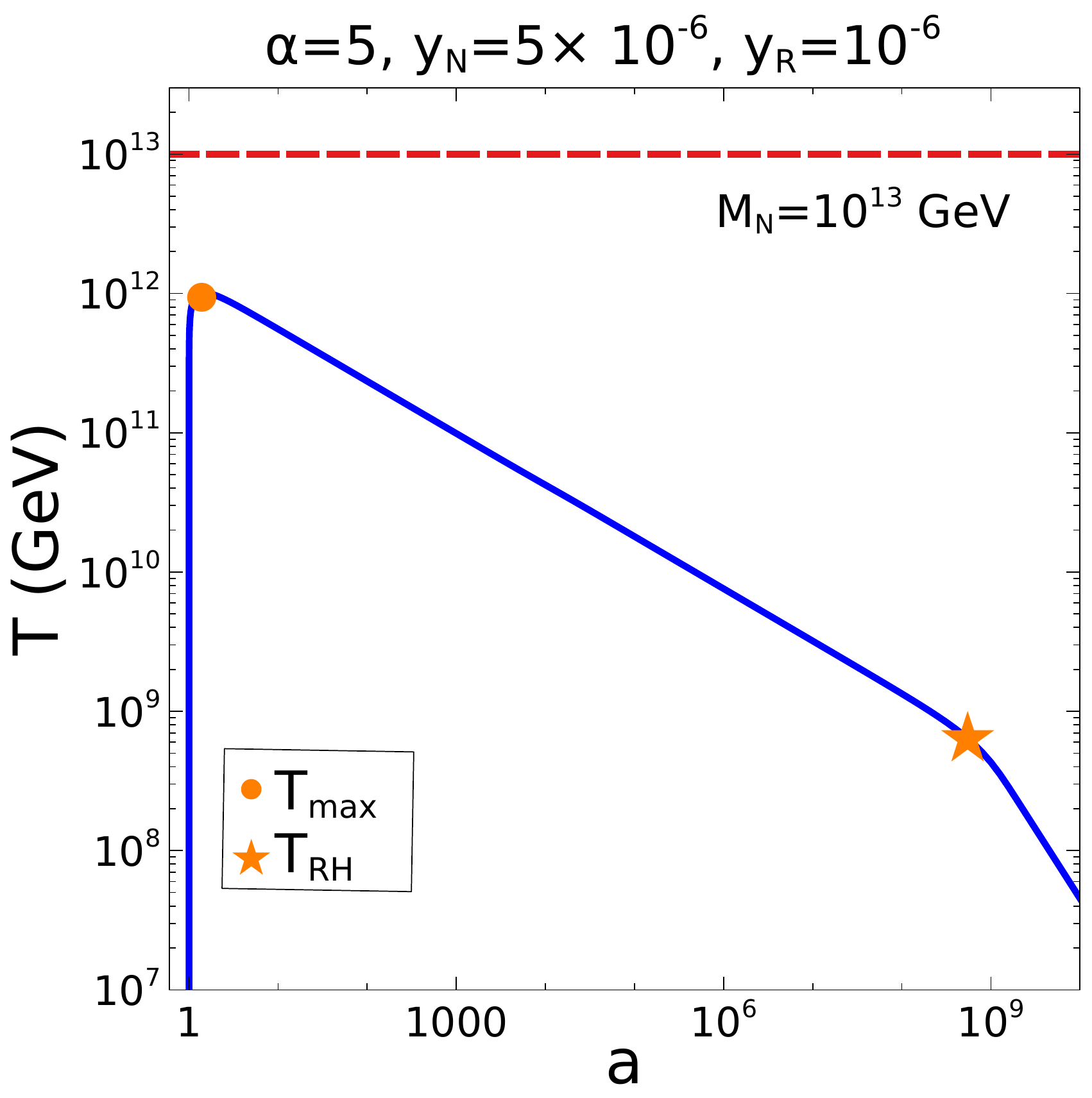}
    \caption{\it Case I: [Left:] Evolutions for energy densities (in $M_P=1$ unit) of different components of the Universe as a function of the scale factor. The yield of the $B-L$ number comoving number density (red curve) with temperature is also shown.
    [Right] The temperature of the Universe is plotted as a function of the scale factor. The red dashed line indicates the RH neutrino mass scale. In both the figures, we have used $M_N=10^{13}$ GeV and $\theta=0.1+0.05i$.}
    \label{fig:LinePlotsCI}
\end{figure*}
\begin{figure*}
    \centering
    \includegraphics[height=7.4cm,width=8.5cm]{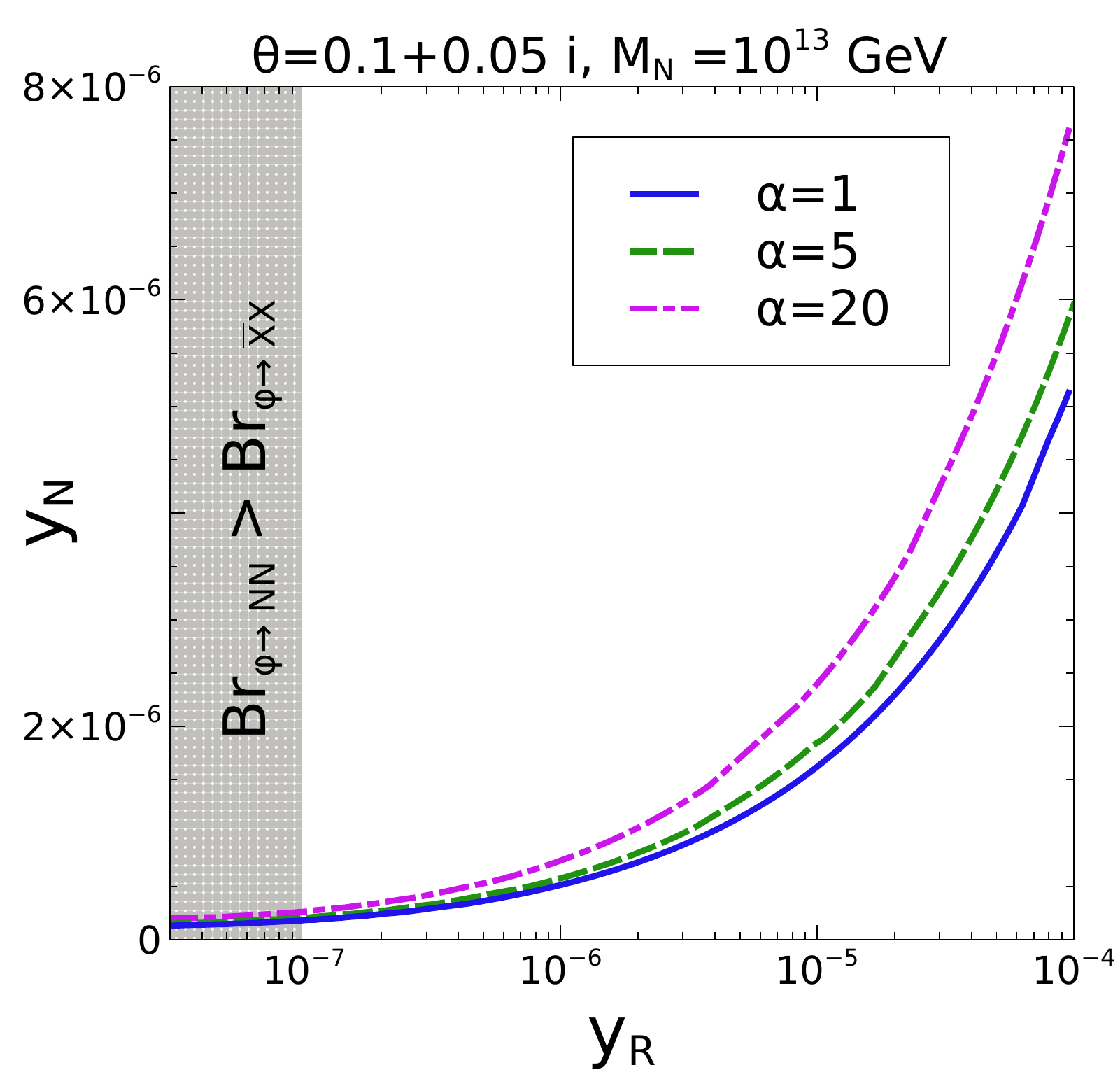}~~
    \includegraphics[height=7.4cm,width=8.5cm]{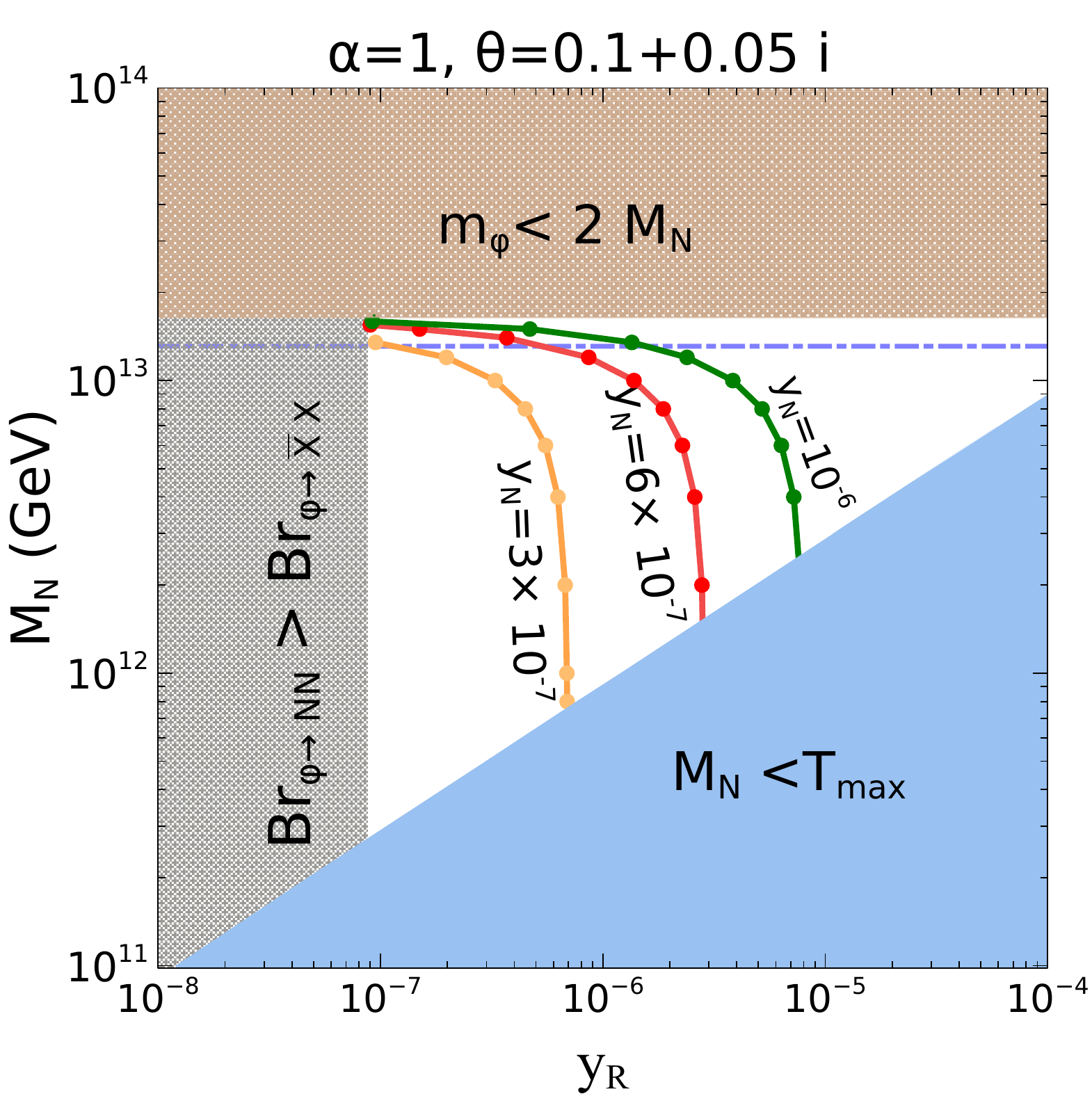}
    \caption{\it Case I: [Left:] Contours for $Y_{B-L}=2.5\times 10^{-10}$ in $y_R-y_N$ plane considering different values of $\alpha$. The shaded region is disfavored for Case I as the inflaton predominantly decays to RHN in this corner of parameter space. [Right:] Allowed region in $y_R-M_N$ plane for $\alpha=1$ after considering all possible constraints: (i) non-thermalisation of RHN, (ii) ${\rm Br}_{\phi\to NN}> {\rm Br}_{\phi\to \overline{X}X}$ and (iii) $m_\phi> 2 M_N$. The blue dashed line indicates the equality $m_\phi=2 M_N$ for $\alpha=20$. Other disfavored colored regions remain almost the same upon varying $\alpha$ in the range 1-20. We also show different $y_N$ lines that give rise to the requisite value of $Y_{B-L}$ for $\alpha=1$.}
    \label{fig:y1y2}
\end{figure*}
\section{Results}
We focus on two possible sub-cases, classified by the interaction strengths between inflaton with radiation and RHNs.  In the first one, the reheating dominantly takes place from the tree-level inflaton decay while in the second case, inflaton decays to RHN first and out-of-equilibrium decay of RHN reheats the Universe. We discuss the results of these two sub-cases separately below:
\begin{figure*}[t]
    \centering
    \includegraphics[height=7.5cm,width=8.8cm]{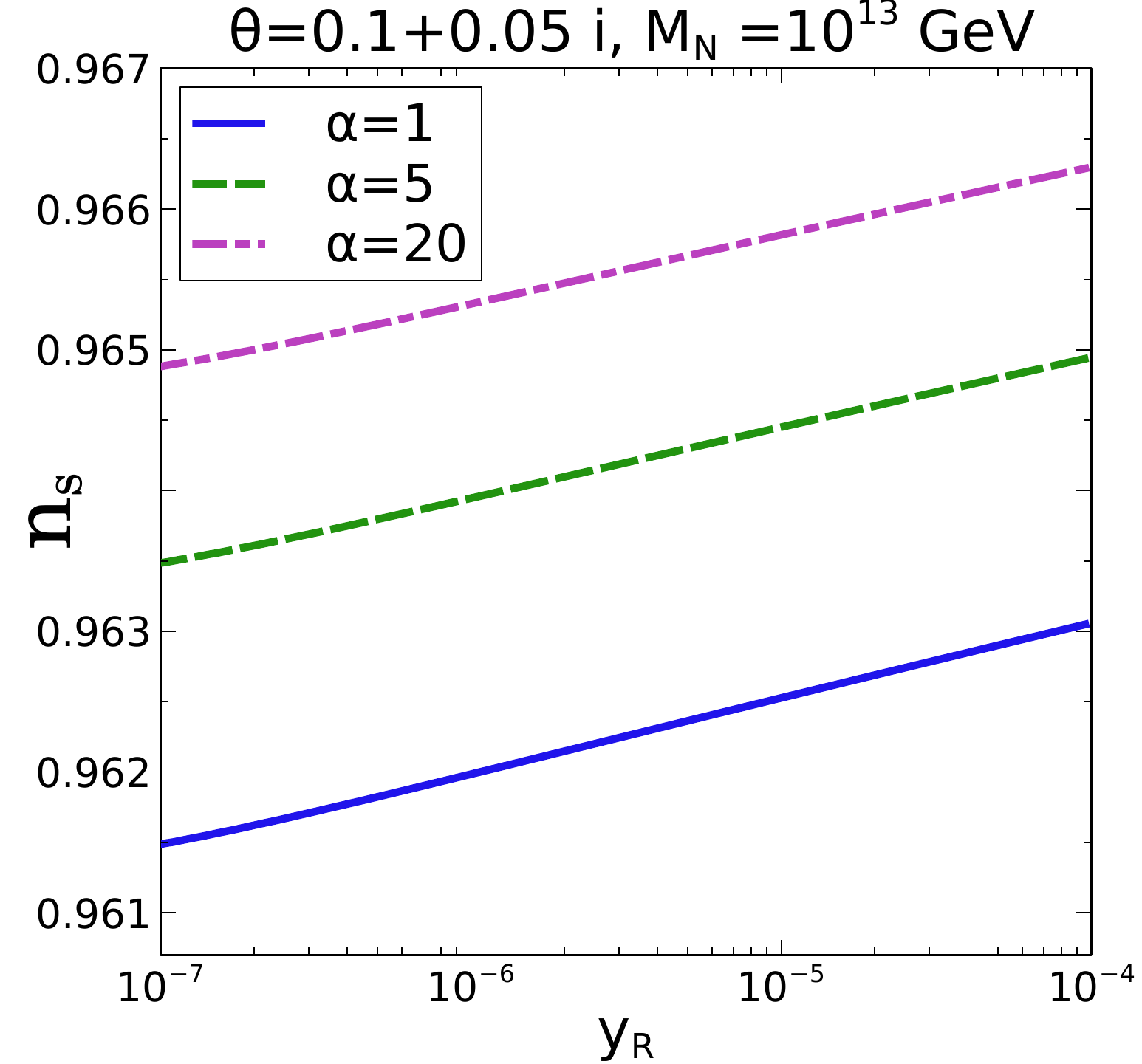}~~~
    \includegraphics[height=7.5cm,width=8.4cm]{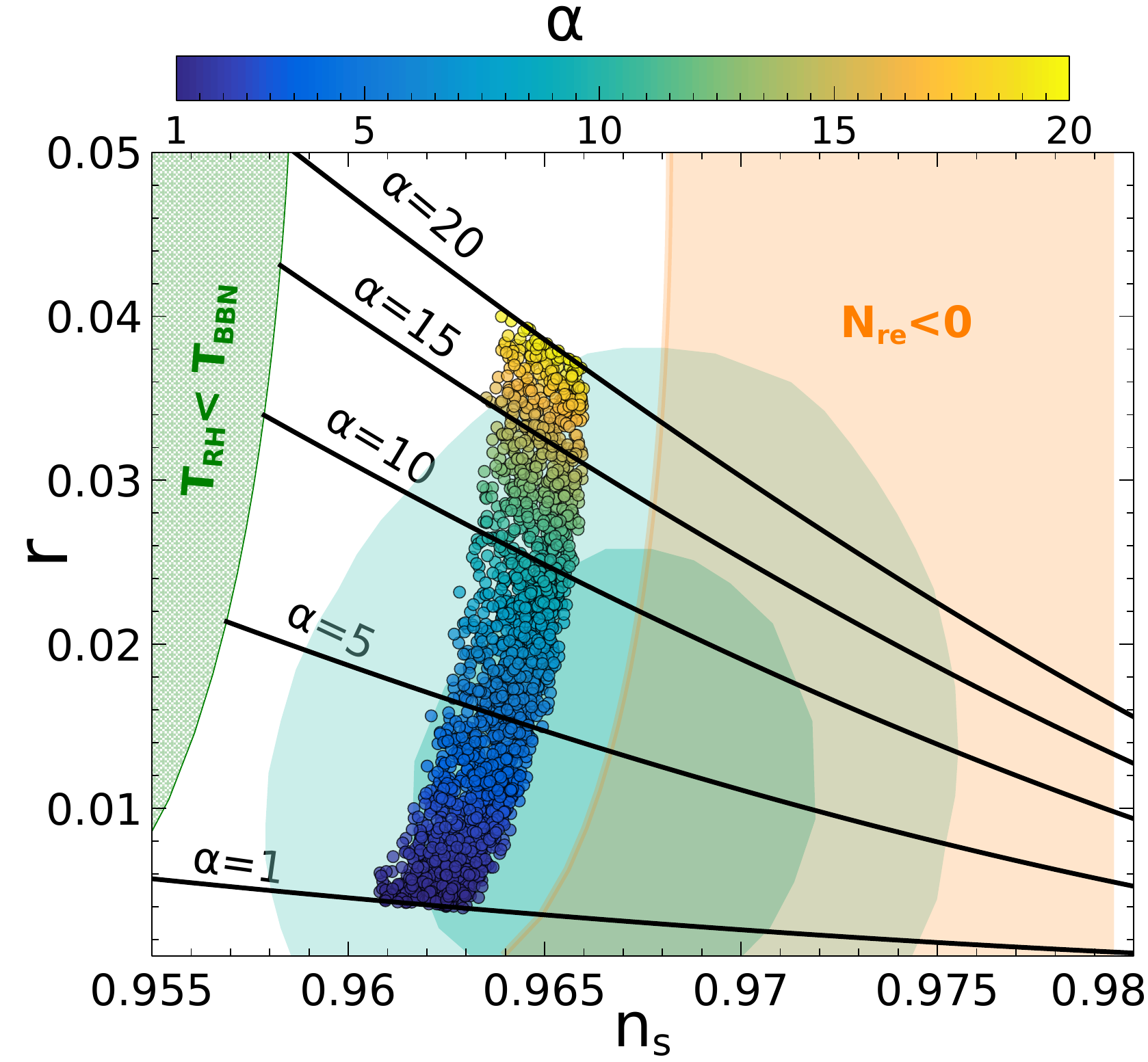}
    \caption{\it Case I: [Left:] Estimate of $n_s$ as a function of $y_R$ for different values of $\alpha$. We fix the range of $y_R$ according to left of Fig.\,\ref{fig:y1y2} that yield the observed amount of baryon asymmetry in the Universe for suitable choices of $y_N$ for a particular value of $\theta=0.1+0.05 i$. [Right:] We show our predictions in the $n_s-r$ plane against the Planck/BICEP $1\sigma$ and $2\sigma$ allowed region \cite{BICEP:2021xfz}. {The five black solid lines represent $(n_s,r)$ predictions corresponding to $\alpha=1,\,5,\,10,\,15$ and 20 from bottom to top respectively for a generic $\alpha-$ attractor inflation model (without non-thermal leptogenesis) where the only constraint is $T_{\rm RH}\gtrsim T_{\rm BBN}\sim 3$\, MeV. The orange shaded region is disfavored due to $N_{\rm re}<0$, since it is unphysical.}
    }
    \label{fig:ns_r}
\end{figure*}

\vspace{5mm}
\noindent\textbf {Case I:} {\color{black}In this category, we have two independent parameters, namely $y_N$ and $y_R$ provided other relevant parameters (including $\alpha$, $\theta$ and $M_N$) are fixed as specified earlier. First in Fig.\,\ref{fig:y2ybl}, we examine the dependence of comoving lepton asymmetry abundance $Y_{B-L}$ ($= \frac{n_{B-L}}{s}$) on the inflaton to radiation coupling coefficient $y_R$ for various choices of $\alpha$. For this purpose, we have fixed $y_N=10^{-6},\theta=0.1+0.05\, i$ and considered $M_N=10^{13}$ GeV (top panel in Fig.\,\ref{fig:y2ybl}) and $M_N=5\times 10^{12}$ GeV (bottom panel in Fig.\,\ref{fig:y2ybl}). We observe that for constant $\alpha$, the $Y_{B-L}$ decreases with the increase of $y_R$. This occurs as a relatively larger $y_R$ leads to a smaller branching ratio of the inflaton decaying into RHNs $(\textrm{Br}_{\phi\rightarrow N N})$ which subsequently results into smaller $B-L$ asymmetry from the RHN decay. In the shaded region of Fig.\,\ref{fig:y2ybl}, the condition of non-thermal leptogenesis $M_N<T_R^{\rm max}$ breaks down and hence disfavored. With a lower RHN mass scale, $M_N>T_R^{\rm max}$ condition rules out an enhanced parameter space compared to the one with a heavier $M_N$. We also highlight the correct range of $Y_{B-L}$ (in lighter red) which can be converted to the observed amount of baryogenesis in the Universe via the sphaleron process before EWPT. The requirement of producing the correct magnitude of $Y_{B-L}$ fixes the allowed value of $y_R$. For example, one finds $y_R\sim 5 \times 10^{-6}$ for $\alpha=5$ with $y_N=1.25\times 10^{-6}$, $\theta=0.1+0.05 i$ and $M_N=10^{13}$ GeV.
It is also noticed that upon increasing $\alpha$, we require a relatively smaller $y_R$ to satisfy the baryon asymmetry bound. This is attributed to the fact that a larger $\alpha$ reduces the mass of the inflaton (since $m_\phi\propto \frac{1}{\sqrt{\alpha}}$) and in turn causes less efficient production of RHN. This results in suppression of the final amount of $Y_{B-L}$. Hence we need a smaller $y_R$ that enhances ${\rm Br}_{\phi\to NN}$ in order to obtain the correct amount of $Y_{B-L}$.} 

We show the evolutions of energy densities of inflaton, radiation and RHN as a function of scale factor in the left panel of Fig.\,\ref{fig:LinePlotsCI} considering a particular benchmark point ($\alpha=5,y_N=5\times10^{-7}$ and $y_R=10^{-6}$ and $\theta=0.1+0.05 i$) that is able to address observed baryon asymmetry of the Universe. It is found that $\rho_N\ll \rho_R$ throughout the 
evolution since in this case, the inflaton predominantly decays to radiation and the RHN is short-lived and never dominates the Universe. Initially, the Universe was $\phi$ dominated and the crossover from $\phi$ to radiation domination is marked by the `$\star$' symbol. In the right panel of Fig.\,\ref{fig:LinePlotsCI}, the evolution of the temperature of the Universe is presented as a function of the scale factor. We also find that the $T_{\rm max}$ turns out to be a few orders larger than the reheating temperature $T_{\rm RH}$ of the Universe as marked by the symbols `$\bullet$' and $`\boldsymbol{\star}$' respectively. We also confirm that the condition for non-thermalisation is easily satisfied since $T_{\rm max}$ remains much below than the RH neutrino mass scale (as indicated by the red dashed line).

\begin{figure*}[htb!]
    \centering
    \includegraphics[height=6.5cm,width=8.2cm]{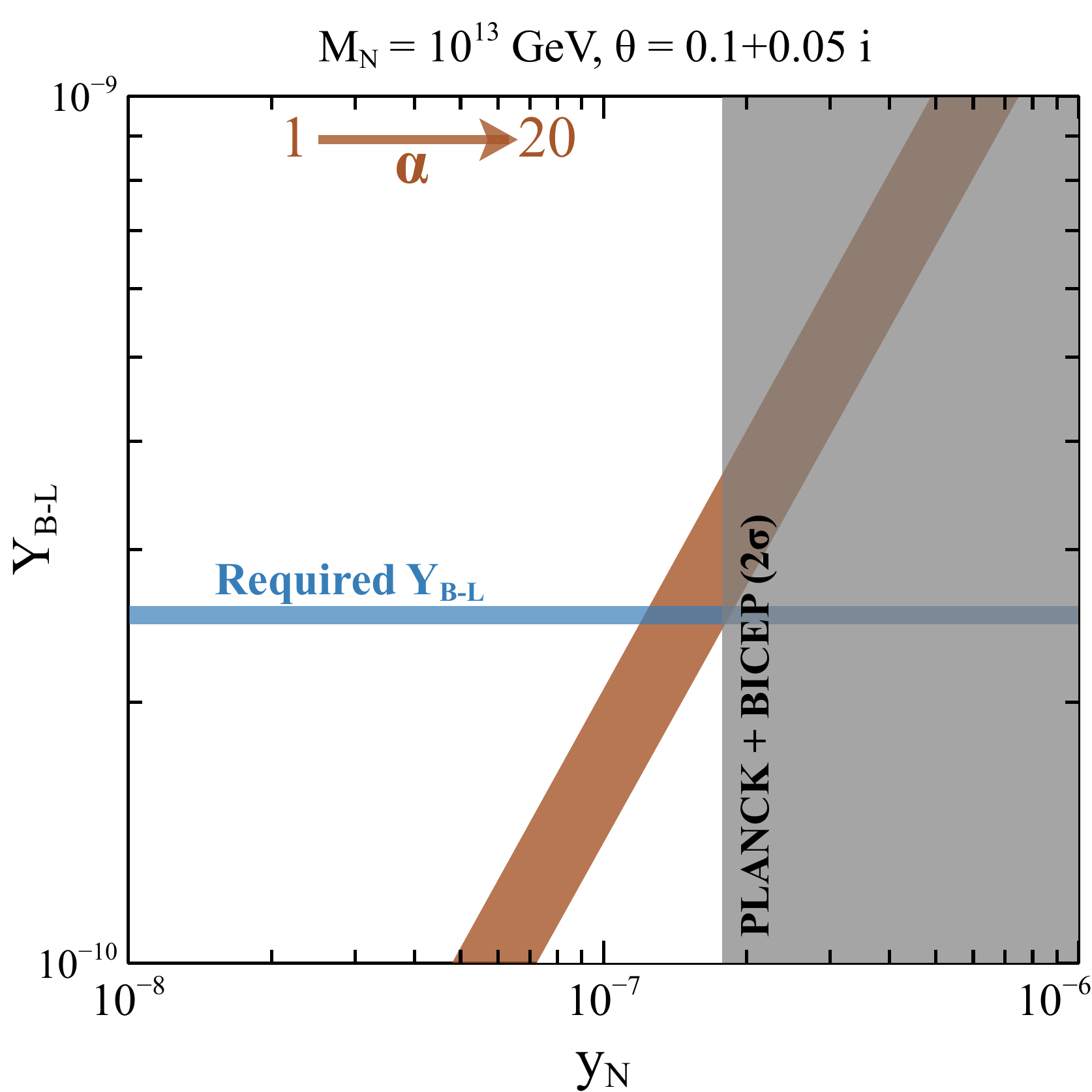}~~~
    \includegraphics[height=6.5cm,width=8.2cm]{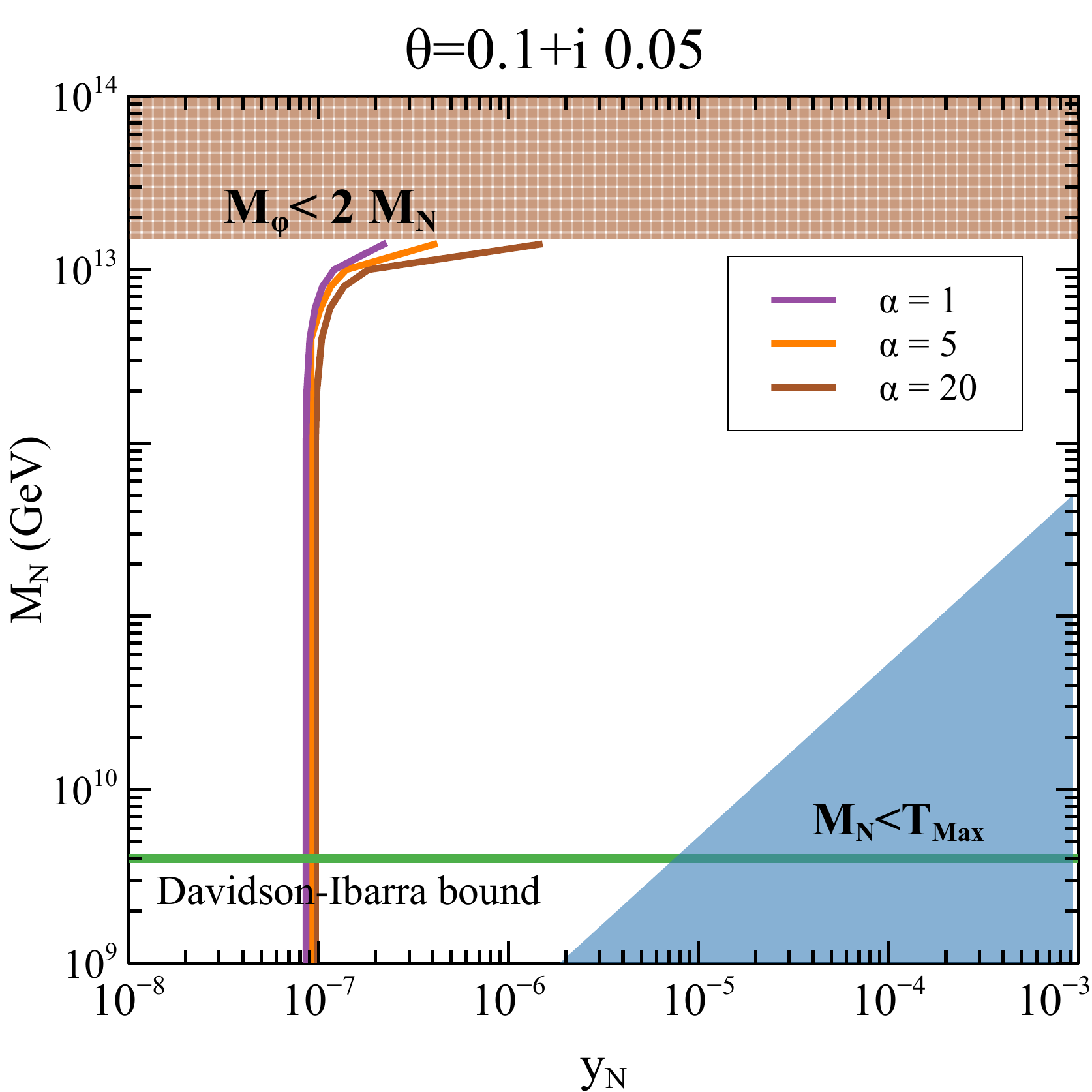}
    \caption{\it Case II: [Left:] The $B-L$ yield is plotted as a function of $y_N$ considering different values of $\alpha$. The shaded region excluding $y_N\lesssim 1.8\times 10^{-7}$ is obtained using the Planck+BICEP $2\sigma$ bound \cite{BICEP:2021xfz}. [Right:] Allowed region in $y_R-M_N$ plane after considering all possible constraints: (i) non-thermalisation of RHN, (ii) $m_\phi> 2 M_N$ and (iii) Davidson Ibara bound \cite{Davidson:2002qv}. The disfavored colored regions remain almost the same upon varying $\alpha$ in the range 1-20 except the upper-bound on $M_N$ changes slightly for a relatively larger $\alpha$ as shown in the right panel of Fig.\,\ref{fig:y1y2}. We also show different $\alpha$ lines that give rise to the requisite value of $Y_{B-L}$. }
    \label{fig:ns_r_case2Ybly1}
\end{figure*}

Next in left of Fig.\,\ref{fig:y1y2} we provide contours of experimentally prefered $Y_{B-L}=2.5\times 10^{-10}$ in the $y_R-y_N$ plane for three different values of $\alpha=(1,5,20)$ and considering $M_N=10^{13}$ GeV and $\theta=0.1+0.05 i$. This figure shows that the required value of $y_N$ increases with the enhancement of $y_R$ to obtain a fixed $Y_{B-L}=2.5\times 10^{-10}$ or {\it vice versa}. As earlier argued, a larger value of $y_R$ reduces the Br$_{\phi\to NN}$ and hence requires larger $y_N$ to attain a particular $Y_{B-L}$. We also observe that a higher value of $\alpha$ favours larger $y_N$ for a fixed $y_R$. As earlier mentioned, the tree level inflaton mass is inversely proportional to $\sqrt{\alpha}$ and hence a larger $\alpha$ means lighter inflaton which corresponds to reduced ${\rm Br}_{\phi\to N N}$ for a particular set of $(y_N,y_R)$. Thus one needs to further tune $y_N$ to a larger value to obtain the same $Y_{B-L}$.
We also point out the region (grey shaded) where ${\rm Br}_{\phi\to N N}>{\rm Br}_{\phi\to \overline{X}X}$ and inflaton decays dominantly to the RHN. We keep $y_R$ below $\mathcal{O}(10^{-4})$ such that $M_N\sim 10^{13}$ GeV remains below $T_{\rm max}$ as evident from top panel of Fig.\,\ref{fig:y2ybl}. Note that the required orders for $y_R$ and $y_N$ ensure the inflationary potential does not receive large radiative corrections arising due to the presence of neutrinos. We infer two important observations from the left panel of Fig.\,\ref{fig:y1y2}. Firstly, the values of $y_R$ and $y_N$ are correlated by the requirement of attaining the correct order of baryon asymmetry. Secondly, we find that the shape of the inflaton potential (represented by varying $\alpha$) leaves some consequences on the estimate of numerical values for $(y_N,y_R)$ to accommodate the correct order of $Y_{B-L}$.

{\color{black}In right panel of Fig.\,\ref{fig:y1y2}, we attempt to find out the allowed range of $M_N$, viable for the non-thermal leptogenesis (baryogenesis) in Case A. Here we have fixed $\alpha=1$ and $\theta=0.1+0.05 i$. In the $y_R-M_N$ plane we impose several constraints namely, (i) $M_{N} > T_{\rm max}$ (blue), (ii) $m_\phi< 2 M_N $ (orange) and (iii) Br$_{\phi\to NN}> {\rm Br}_{\phi\to\overline{X} X}$. We have also shown the different $y_N$ lines that yield observed values of $Y_{B-L}$ considering $\alpha=1$. We find $2.2\times 10^{11}{\rm \, GeV}\lesssim M_N\lesssim 1.6\times 10^{13}{\rm \,GeV}$ as a viable range for $M_N$ suitable for non-thermal leptogenesis in Case I. {Importantly,  this bound on $M_N$ is more or less independent of the CP violating parameter $\theta$ since none of the three above-mentioned constraints depends on $\theta$.} The upper-bound on $M_N$ gets slightly stronger for a larger $\alpha(\leq 20)$ as indicated by blue dot-dashed line. Since $m_\phi\propto \frac{1}{\sqrt{\alpha}}$, the $m_\phi> 2 M_N$ bound gets violated at a relatively smaller $M_N$ for a larger $\alpha$. In contrast, the lower bound on $M_N$ remains more or less the same for a different upon varying $\alpha(\leq 20)$. One also finds from the right panel of Fig.\,\ref{fig:y2ybl} that for each $y_N$ line the requisite value of $y_R$ is almost insensitive to the change of $M_N$ beyond some critical value of the same quantity $M_N$. As an example, with $y_N=3\times 10^{-7}$, the obtained value of $y_R$ remains almost constant for varying $M_N$ below $6\times 10^{12}$ GeV. In the preceding text, we will discuss the impact of the allowed parameter space, pertaining to non-thermal leptogenesis on the predictions for spectral index and tensor to scalar ratio of the $\alpha$-attractor inflation model.}

In Fig.\,\ref{fig:ns_r}, we depict the observational consequence of successfully producing the right order of baryon asymmetry in the measurements of the inflationary observables namely $n_s$ and $r$. In the left panel, we utilize the allowed ranges for $y_R$ from Fig.\,\ref{fig:y1y2} for different $\alpha$ with $M_N=10^{13}$ GeV and $\theta=0.1+0.05 i$ in order to satisfy the baryon asymmetry of the Universe and find the predictions for spectral index $n_s$. Remember for each $\alpha$, a particular $y_R$ corresponds to a definite value of $y_N$ in Fig.\,\ref{fig:y1y2}. Increasing $y_R$ leads to smaller $N_{\rm re} (> 0)$ according to Eq.(\ref{eq:Nre}) which translates into bigger $n_s$. 

{Up to this, we have kept the CP-violating parameter $\theta$ fixed to a specific value. In order to provide a concrete and generic (to be verified later) predictions for $n_s$ and $r$ in the presence of non-thermal leptogenesis, we perform a numerical scan in the range $1\leq \alpha\leq 20$ and $2.2\times 10^{11} {\rm\, GeV}\lesssim M_N\lesssim M_N^{\rm max}$ by varying $\theta$ randomly. Note that $M_N^{\rm max}$ is the maximum allowed value of $M_N$ as a function of $\alpha$. For example, $M_N^{\rm max}\simeq 1.6\times 10^{13}$ GeV for $\alpha=1$ and $\theta=0.01+0.05 i$ (see right panel of Fig.\,\ref{fig:y1y2}). 
While obtaining the predictions for $n_s$ and $r$, we impose five relevant selection criterion: (i) adequate yield of $\eta_b$, (ii) decay of $N$ gets completed before electroweak (EW) phase transition temperature, (iii) $M_N< T_{\rm max}$, (iv) Br$_{\phi\to NN}< {\rm Br}_{\phi\to X X}$ and finally (v) $Y_\nu^2\lesssim 0.5$ \cite{Bambhaniya:2016rbb} to ensure that EW vacuum does not turn unstable.
The final estimate of $n_s$ and $r$ (giving rise to the correct value of $Y_{B-L}$) are presented in the right panel of Fig.\,\ref{fig:ns_r} along with the Planck/BICEP allowed $1\sigma$ and $2\sigma$ contours \cite{BICEP:2021xfz}. It is noticed that for a constant value of $\alpha$, the predicted ranges for tensor to scalar ratio ($r$) and $n_s$ are strictly restricted. In fact, the requirement of successful baryogenesis indeed constraints the $n_s-r$ plane which is more stringent than the one in a generic $\alpha-$attractor inflation model (without non-thermal leptogenesis), subject to a single constraint, $T_{\rm RH}\gtrsim T_{\rm BBN}\sim 3$\, MeV that ensures the success of BBN does not get altered.} We also note that the predicted region of ($n_s-r$) almost remains inside the 2$\sigma$ sensitivity curve of Planck/BICEP data \cite{BICEP:2021xfz} (except a small portion of the $18\leq \alpha\leq 20$ region). We infer that future CMB experiments such as  CBM-S4, STPOl, LitBird ~\cite{CMB-S4:2020lpa,Hazumi:2019lys,Adak:2021lbu,SPT:2019nip,POLARBEAR:2015ixw,ACT:2020gnv,Harrington:2016jrz,LSPE:2020uos, Mennella:2019cwk, SimonsObservatory:2018koc,SPIDER:2021ncy,Sehgal:2019ewc,CMB-HD:2022bsz,CMB-HD:2022bsz}) with improved sensitivities or more precise measurements of $(n_s,r)$ may validate/refute the relevance of non-thermal leptogenesis in $\alpha$-attractor inflation models.

{Till now, we do not know whether the prediction for $n_s$ and $r$ in case I is generic or not. By generic, we mean the outcomes of $n_s$ and $r$ are independent of whether a radiation dominated Universe is obtained from inflaton or RHN decay. To check this fact, we now proceed to case II, where inflaton decaying to radiation is substantially suppressed and a radiation dominated Universe is obtained purely from the RHN decay.}

\vspace{5mm}
\noindent
\textbf{Case II:}
\begin{figure}[h]
    \centering
   \includegraphics[height=6.5cm,width=8cm]{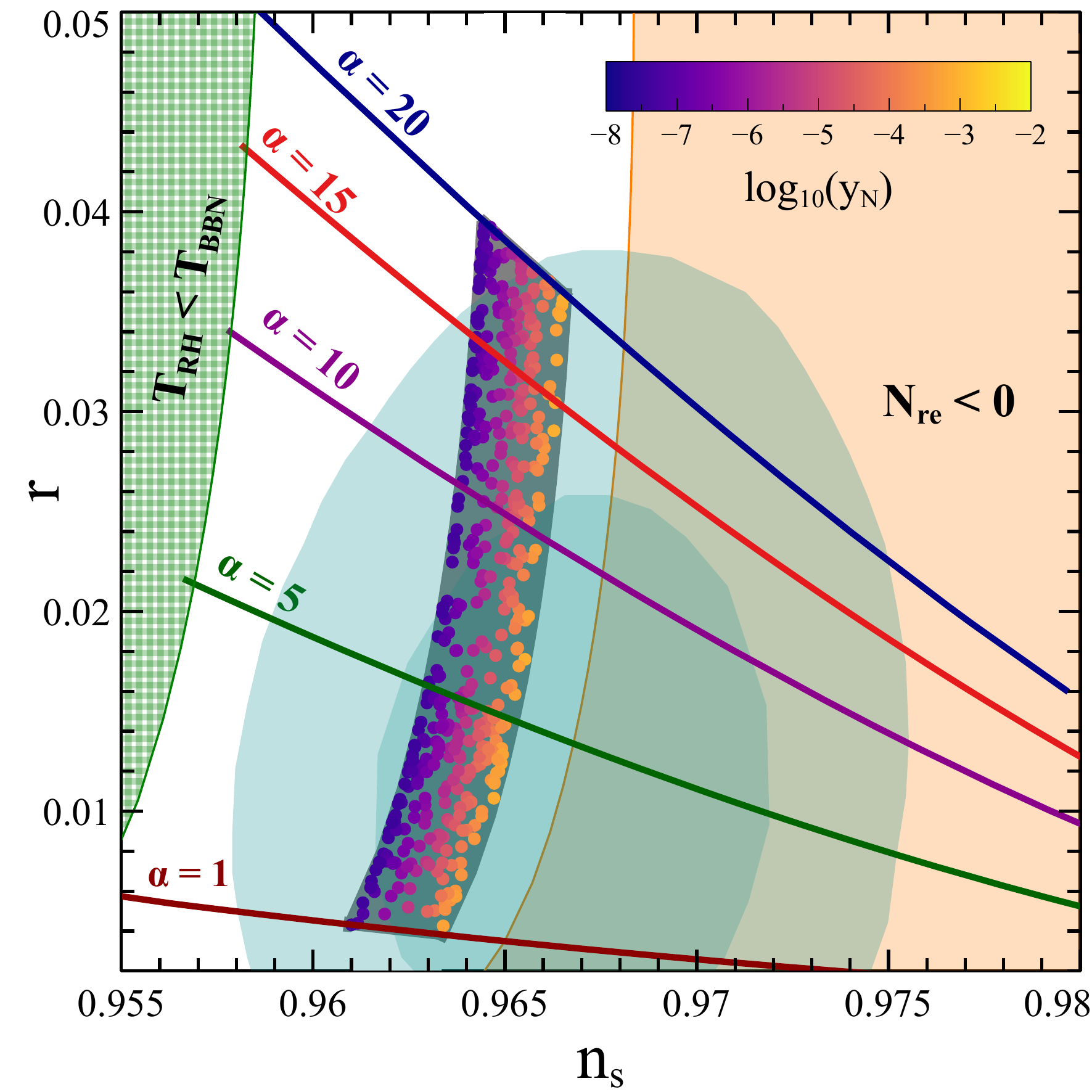}
    \caption{\it Case II: With the values of $y_N$ obtained for different $\alpha$ and $M_N$ that give rise to observe baryon asymmetry, we find the corresponding predictions for $n_s$ and $r$. We also present the $1\sigma$ and $2\sigma$ allowed regions by Planck/BICEP \cite{BICEP:2021xfz} in the same plane. {The five black solid lines represent $(n_s,r)$ predictions corresponding to $\alpha=1,\,5,\,10,\,15$ and 20 from bottom to top respectively for a generic $\alpha-$ attractor inflation model (without non-thermal leptogenesis) where the only constraint is $T_{\rm RH}\gtrsim T_{\rm BBN}\sim 3$\, MeV. The orange shaded region is disfavored due to $N_{\rm re}<0$, since it is unphysical.}}
    \label{fig:ns_r_case2nsr}
\end{figure}
In this case, both the lepton (baryon) asymmetry and reheating of the Universe are controlled by $y_N$ since we have considered here $y_R\simeq 0$. The inflaton decays to RHN first and its subsequent out-of-equilibrium decay reheats the Universe and simultaneously yields the lepton asymmetry. Therefore, in this case, the reheating of the Universe is a two-step process. Once again, we have solved the set of Boltzmann equations (Eqs.(\ref{eq:leprhoPh} - \ref{eq:lepAs})) numerically considering $\Gamma_\phi^R\simeq 0$. First in the left panel of Fig.\,\ref{fig:ns_r_case2Ybly1}, we have shown the evolution of $Y_{B-L}$ as a function of $y_N$ by varying $\alpha$ in the range $1\leq \alpha\leq 20$. Here we have set $M_N=10^{13}$ GeV and $\theta=0.1+0.05 i$. For a fixed $\alpha$, we find $Y_{B-L}$ to increase with $y_N$. This happens since an increasing $y_N$ enhances the rate of RHN production as well as the reheating temperature which results into larger $Y_{B-L}$. On the other hand, a larger $\alpha$ reduces the inflaton mass which in turn suppresses the RHN production rate, thus reducing the lepton asymmetry yield. To compensate for that, a larger $y_N$ is required for a larger value of $\alpha$. We also notice that satisfaction of Planck 2$\sigma$ bound on the $n_s-r$ plane (see Fig.\,\ref{fig:ns_r_case2nsr}) restricts $y_N\lesssim 1.8\times 10^{-7}$. Therefore future CMB experiments, CMB-S4, SPTpol, LitBIRD and CMB-Bharat with upgraded sensitivities have the capability to rule out the complete parameter space favoring non-thermal leptogeneis. In the right panel of Fig.\,\ref{fig:ns_r_case2Ybly1}, we attempt to identify the allowed region in $y_N-M_N$ plane that is suitable for non-thermal leptogenesis in the present case. We have imposed the relevant conditions, (i) non-thermalisation of RHN ($M_N>T_{\rm max}$) at early Universe, (ii) $m_\phi< 2M_N$ that prohibits the tree level inflaton decay to RHN and (iii) Davidson Ibara bound \cite{Davidson:2002qv}. We find that in case II, non-thermal leptogenesis is possible in the range of $4\times 10^{9}{\rm \, GeV}\lesssim M_N\lesssim \mathcal{O}(10^{13}) \,\text{GeV}$. This allowed range of $M_N$ remains more or less the same upon varying $\alpha$ and $\theta$, similar to the one in the previous case. We have also shown different $\alpha$ lines that provide the correct amount of baryon asymmetry in the same plane. Similar to the previous case, we also notice here the requisite value of $y_N$ remains almost constant upon varying $M_N$ below $10^{13}$ GeV.

{Finally in Fig.\,\ref{fig:ns_r_case2nsr}, we show the $n_s-r$ predictions for Case II within the allowed ranges of $M_N$ for different values of $\theta\in\{-4\pi,4\pi\}$ as procured in right panel of Fig.\,\ref{fig:ns_r_case2Ybly1}. We use the estimate of $y_N$ for a particular $\alpha$ and $M_N$ that gives rise to correct $Y_{B-L}$ to compute $r$. Once again we impose a few relevant conditions such as, (i) adequate yield of $\eta_b$, (ii) decay of $N$ gets completed before electroweak (EW) phase transition temperature, (iii) $M_N< T_{\rm max}$, and (iv) $Y_\nu^2\lesssim 0.5$ \cite{Bambhaniya:2016rbb} to ensure that EW vacuum does not turn unstable. We also highlight the regions that are disfavored by $T_{\rm RH}<T_{\rm BBN}$ and $N_{\rm re}<0$. The predictions for $n_s-r$ is exactly similar to the case I, favoring non-thermal leptogenesis, much more stringent than the one in a general $\alpha-$attractor inflation model. This is expected since the connection between non-thermal leptogenesis and predictions for $(n_s,r)$ is established by the preferred range of reheating temperature,  depending on the value of lepton asymmetry parameter $\varepsilon$ and inflaton mass.}

\section{A semi-analytical approach} 
The computation of $n_s$ and $r$ in the previous section is performed numerically in a detailed manner. What we found is that the values of $n_s$ and $r$ are strictly restricted to ensure a consistent picture for non-thermal leptogenesis,  compared to a generic $\alpha-$attractor inflation model. The reason lies in the fact the reheating temperature of the Universe can't be arbitrary once we consider non-thermal leptogenesis. This feature can be understood in a simple manner by a semi-analytic calculation. The lepton asymmetry produced from the decay of RHN can be analytically approximated as (under the approximation of instantaneous decay for both the inflaton and RHN) \cite{Hamaguchi:2002vc,Fukuyama:2005us},
\begin{align}
    \frac{n_B}{s}\,\approx \,& 8.7 \times 10^{-11} \text{Br}_{\phi\rightarrow NN} \left(\frac{T_R}{1.95\times10^{6} \text{ GeV}} \right)\nonumber\\ & ~~~~~~~~~~~~~~~~~\times \left(\frac{2\,M_{N}}{M_{\phi}} \right)\left(\frac{m_{\nu_3}}{0.05 \text{ eV}}  \right)\delta_{\rm eff},
\end{align}
 where $\delta_{\rm eff}$ is the effective CP violation. 
 Now, in order to have observed amount of $\frac{n_B}{s}\sim 8.7\times 10^{-11}$, the minimum value of reheating temperature would be,
\begin{align}\label{TRmin}
 T_{R}^{\rm min}\simeq 1.95\times 10^{6}\,\text{GeV}. 
\end{align}
considering all other relevant parameters are set at their respective maximum values including maximal CP violation and $m_{\nu_3}\simeq 0.05$ eV.
 
 On the other hand, the condition of non-thermal leptogenesis dictates $M_N> T_{\rm max}$. Now, $M_N\leq \frac{m_\phi}{2}$ implies $T_{\rm max}<\frac{m_\phi}{2}$. The value of maximum reheating temperature, $T_R^{\rm max}$ is connected to maximum temperature of the Universe by the following relation \cite{Chung:1998rq},
 \begin{align}
\frac{T_{\rm max}}{T_{\rm R}^{\rm max}}=0.77\left(\frac{9}{5\pi^3 g_*}\right)^{1/8}\left(\frac{\sqrt{8\pi}H_kM_P}{T_{R}^{\max^2}}\right)^{1/4},     
 \end{align}
 where $g_*\,\simeq\,106.25$ is the number of energy relativistic degrees of freedom. Subsequently, we find (using $H_k=\frac{\pi M_P^2 r A_s^{\rm obs}}{\sqrt{2}}$ and Eq.(\ref{eq:r})),
 \begin{align}\label{TRmax}
 T_{R}^{\rm max}\simeq 6.25\times 10^{-8}M_P\left(\frac{5 \pi ^3 g_{*\rm SM}}{9}\right)^{1/4} \frac{(y-1)^{3/2} (y+3)^2}{\alpha ^{7/4}}
 \end{align}
where $y^2=1+3\alpha(1-n_s)$. Considering instantaneous reheating, we can write $\rho_{\rm re}=\frac{\pi^2}{30}g_*T_R^4$. Then Eq.(\ref{eq:Nre-inf}), the governing equation to determine $n_s$ is translated to, 
\begin{widetext}
\begin{align}\label{eq:nsana}
& 61.573 + 0.86 \sqrt{\alpha }-\frac{3 \alpha }{y-1}+0.75\, \alpha  \log \left(\frac{y+3}{y-1}\right)-0.75 \,\alpha  \log \left(\frac{2}{\sqrt{3\,\alpha} }+1\right)\nonumber\\
&+\frac{1}{2} \log \left(\frac{0.00024 \left(\sqrt{\alpha }+1.154\right) (y-1)^2}{\sqrt{\alpha } \left(y^2+2 y-3\right)}\right)
+0.083 \log \left(\frac{7.256\times 10^9 \left(\sqrt{\alpha }+1.154\right)^2 \,\alpha\,  T_R^4}{M_P^4 (y-1)^2 (y+3)^2}\right)\,=\,0\,.
\end{align}
\end{widetext}
The L.H.S of Eq.(\ref{eq:nsana}) is a non-linear function of $n_s$ and thus Eq.(\ref{eq:nsana}) correlates $n_s$ with $T_R$. We use the minimum and maximum permitted values of $T_R$ as demonstrated in Eqs.(\ref{TRmin}) and (\ref{TRmax}) respectively and subsequently find the preferred ranges for $n_s$ (using Eq.(\ref{eq:nsana})) corresponding to different $\alpha$ values. From Table\,\ref{tab:NTM-nsr}, it is pretty clear that the results found from numerical analysis are almost overlapping with the ones, computed in the semi-analytic way. This reinforces our previously drawn conclusion that occurrence of non-thermal leptogenesis to yield observed baryon asymmetry restrict the reheating temperature of the Universe in a certain range ($T_R^{\rm min}<T_R<T_R^{\rm max}$) and that in turn set the lower and upper limits for $n_s$. We have also checked that the preferred range of $r$ corresponding to a particular $\alpha$ also remain more or less consistent with our numerical findings.

\begin{table}[h]
\renewcommand*{\arraystretch}{1.4}
    \centering
    \begin{tabular}{| c | c | c |}
    \hline
         $~\alpha~ $ & ~$n_s$ (numerical)~ & ~$n_s$ (semi-analytical)~  \\ \hline
          1 & 0.9607\,-\,0.9631 & 0.9606\,-\,0.9633\\ \hline
          5 & 0.9625\,-\,0.9651 & 0.9626\,-\,0.9651\\ \hline
          10 & 0.9632\,-\,0.9657 & 0.9634\,-\,0.9658\\ \hline
          15 & 0.9635\,-\,0.9661 & 0.9638\,-\,0.9662\\ \hline
          20 & 0.9638\,-\,0.9661 & 0.9640\,-\,0.9663\\ \hline
    \end{tabular}
    \caption{Favored ranges for $n_s$, corresponding to different $\alpha$ values that give rise to a consistent scenario of non-thermal leptogenesis in the ealry Universe. We provided the results obtained both numerically and semi-analytically for comparison purpose.}
    \label{tab:NTM-nsr}
\end{table}

\section{Conclusion}
We re-analyse the non-thermal production of matter-antimatter asymmetry in the early universe via leptogenesis and investigated its impact on CMB predictions for inflationary observables namely spectral index and tensor to scalar ratio. Considering very massive RHN, we reinforce that the final amount of lepton asymmetry yield crucially depends on the reheating dynamics of the Universe. Consequently, we find that such correlations result into very predictive inflationary observable values ($n_s,r$). Although for concrete predictions we considered the $\alpha$-attractor inflationary model for our exercise, the prescription presented here is generic and can be implemented in any inflationary scenario. We scrutinized two possible sub-cases: (i) inflaton decays dominantly to radiation and (ii) inflaton does not couple to radiation directly but a standard radiation-dominated Universe is secured via inflaton decaying to heavy neutrinos and subsequently to radiation. Below we summarise the important observations that came out from our analysis.
{\begin{itemize}
    \item The RH mass scale is restricted for both case I and case II to realize successful baryogenesis through non-thermal leptogenesis. In case I, the suitable range for non-thermal leptogenesis is $2.2\times 10^{11}{\rm \, GeV}\lesssim M_N\lesssim \mathcal{O}(10^{13})$\, GeV. For case II, the allowed range for the same is $4\times 10^{9}{\rm \, GeV}\lesssim M_N\lesssim \mathcal{O}(10^{13})$\, GeV.  The bound on the RHN mass scale in both cases is independent of the CP violating parameter $\theta$. In case I, the bound on $M_N$ gets stringent for larger inflaton-radiation coupling strength whereas in case II the same happens for larger inflaton-RHN coupling strength.
    \item Corresponding to a fixed value of $\theta$ and $M_N$, Planck 2$\sigma$ data can constrain the inflaton-RHN coupling in case II, giving rise to sufficient $\eta_B$. For example in Fig.\,\ref{fig:ns_r_case2Ybly1}, we have  obtained $y_N\lesssim 1.8\times 10^{-7}$ for $M_N\sim 10^{13}${\rm \,GeV} and $\theta=0.1+0.05 i$. Consideration of Planck 1$\sigma$ bound will further restrict the allowed range of $y_N$.

    \item  Crucially, we have determined the estimates of $n_s$ and $r$ in both cases, resulting in a consistent non-thermal leptogenesis scenario and the outcomes are more or less identical. This result is anticipated since the amount of baryon asymmetry solely depends only on the lepton asymmetry parameter and reheating temperature, not on how a standard radiation \footnote{By standard Universe we mean $\rho_N\ll {\rm max}[\rho_\phi,\rho_R]$ always.} dominated Universe is ensured in the post inflationary epoch. And the same amount of reheating temperature also governs the period of reheating ($N_{\rm re}$) which subsequently fixes $n_s$ and $r$. For reader's convenience in table\,\ref{tab:NTM-nsr} we provide the generic allowed range for $n_s$ (obtained from both numerical and semi-analytical anlyses), corresponding to a few $\alpha$ values for the successful occurrence of non-thermal leptogenesis in the early Universe. With this, we infer that in order to ensure a successful non-thermal leptogenesis in a $\alpha$-attractor inflation model, the predictions for $n_s$ must fall inside the projected ranges as mentioned in table\,\ref{tab:NTM-nsr}, corresponding to different $\alpha$ values.
\end{itemize}

In summary, we have demonstrated that a consistent non-thermal leptogenesis scenario indeed provides a detectable imprint in the CM predictions for $n_s$ and $r$ which perhaps can be utilised as a scope to indirectly test high-scale non-thermal leptogenesis. To the best of our knowledge, such a link between the successful occurrence of non-thermal leptogenesis in the early Universe and $n_s,r$ prediction has been put forward for the first time. Future CMB experiments with improved sensitivities like LitBIRD, SPTpol, CMB-S4, CMB-HD, CMB-Bharat and others~\cite{CMB-S4:2020lpa,Hazumi:2019lys,Adak:2021lbu,SPT:2019nip,POLARBEAR:2015ixw,ACT:2020gnv,Harrington:2016jrz,LSPE:2020uos, Mennella:2019cwk, SimonsObservatory:2018koc,SPIDER:2021ncy,Sehgal:2019ewc,CMB-HD:2022bsz,CMB-HD:2022bsz}) will be able to eliminate part of the parameter space further that gives rise to consistent non-thermal leptogenesis (baryogenesis) scenario.}

\section{Acknowledgements}
We gratefully acknowledge the anonymous referee for the valuable suggestions and recommendations. DN would like to thank Arghyajit Datta for various discussions. 
AKS is supported by NPDF grant PDF/2020/000797  from Science and Engineering Research Board, Government of India. AKS also acknowledges the Institute of Physics, Bhubaneswar for a postdoctoral fellowship. The work of DN is supported by the National Research Foundation of Korea (NRF)’s grants, grant no. 2019R1A2C3005009(DN).

\appendix
\section{Casas-Ibara parameterization}\label{sec:CI}
 The $\nu$MSM model \cite{Asaka:2005an} with two Majorana RH neutrinos ($N_{1,2}$) can successfully address the neutrino oscillation data where the lightest SM neutrino mass vanishes. If one of the RH neutrinos turns heavier than the inflaton mass, it does not contribute to the production of lepton asymmetry. Hence, considering the mass spectrum $M_{2}\gg m_\phi>M_1$, the $N_1$ can be produced efficiently in the early Universe from inflaton decay. In a subsequent process, the decay of $N_1$ yields lepton asymmetry.

The CP asymmetry produced from the decay of the lightest right handed neutrinos can be written as \cite{Covi:1996wh},
\begin{equation}
    \varepsilon=\frac{1}{8\pi}\sum_{j\neq 1} \frac{\text{Im}\left[ \left(Y_\nu^\dagger Y_\nu \right)^2_{1j}\right]}{\left(Y_\nu^\dagger Y_\nu \right)_{11}}\mathcal{F}\left( \frac{M_j^2}{M_1^2}\right)
\end{equation}
where $\mathcal{F}(x)\approx -3/2\sqrt{x}$. Adopting the familiar Casas-Ibara parameterization the Yukawa matrix $Y_\nu$ can be written as \cite{Casas:2001sr}
\begin{equation}
    Y_\nu = \frac{\sqrt{2}}{v} U_{\rm {PMNS}}\sqrt{m_\nu^d}\mathcal{R}^T\sqrt{M_N}
\end{equation}
where $v=246$ GeV and $\mathcal{R}$ is a $3\times 3$ orthogonal matrix and is chosen as (for normal hierarchy) \cite{Antusch:2011nz}, 
\begin{equation}
 \mathcal{R}=   \left(
\begin{array}{ccc}
 0 & \cos \theta  &  \sin \theta  \\
 0 & - \sin\theta  & \cos \theta  \\
 1 & 0 & 0 \\
\end{array}
\right),
\end{equation}
where $\theta$ is a complex angle.
Using the best-fit values of the neutrino oscillation data \cite{ParticleDataGroup:2022pth} in the massless limit of for the lightest active neutrino mass, we obtain,
\begin{align}
\varepsilon=\frac{3}{16\pi}\times C\times \left(\frac{M_1}{v}\right){\rm Im}\left[\frac{\left\{A \cos\theta^*\sin\theta-\cos\theta\sin\theta^*\right\}^2}{-A\cos\theta^*\cos\theta-\sin\theta^*\sin\theta}\right], 
\label{eq:assyP}
\end{align}
where,
$ C=2.40\times 10^{-12}$ and $A=5.81$ are the numerical constants. 
The decay width of the $N_1$ takes the following form:
\begin{align}
    \Gamma_{N_1}=\frac{M_1}{16\pi}\times\left(\frac{M_1}{v}\right)\times L\times \left(\cos\theta\cos\theta^*+\frac{29}{5}\sin\theta\sin\theta^*\right),
\end{align}
where $L=2.94\times 10^{-32}$ is another numerical constant.
\bibliography{ref}
\end{document}